Title page

Original article

# POINT: a web-based platform for pharmacological investigation enhanced by multi-omics networks and knowledge graphs

Zihao He[a,b,c,†], Liu Liu[c,†], Dongchen Han[c,†], Kai Gao[a,b,c], Lei Dong[c], Dechao Bu[b], Peipei Huo[b], Zhihao Wang[b], Wenxin Deng[c], Jingjia Liu[a], Jin-cheng Guo[c,*], Yi Zhao[b,*], Yang Wu[b,*]

[a] *Ningbo No.2 Hospital, Ningbo 315010, China*

[b] *Research Center for Ubiquitous Computing Systems, Institute of Computing Technology, Chinese Academy of Sciences, Beijing, 100190, China*

[c] *School of Traditional Chinese Medicine, Beijing University of Chinese Medicine, Beijing, 100029, China*

[†] The first three authors should be regarded as Joint First Authors.

[*] Corresponding author. Tel: +86 10 6260 0822; Fax: +86 10 6260 1356.

Email addresses: wuyang@ict.ac.cn (Yang Wu), biozy@ict.ac.cn (Yi Zhao), guojincheng@bucm.edu.cn (Jincheng Guo)

**Funding**

This work was supported by the National Key R&D Program of China [2021YFC2500203]; the National Natural Science Foundation of China [32070670, 32341019]; Beijing Natural Science Foundation Haidian Origination and Innovation Joint Fund [L222007]; Ningbo Major Project for High-Level Medical and Healthcare Teams [2023030615]; Ningbo Science and Technology Innovation Yongjiang 2035 Project [2024Z229]; Major Project of Guangzhou National Laboratory [GZNL2023A03001]; Beijing Nova Program [20240484661]; Funding for open access charge: National Key Research and Development Program of China.

**Author contributions**

ZiHao He: Conceptualization; Data analysis; Methodology; Writing - original draft. Liu Liu, Dongchen Han and Kai Gao: Resources; Data curation. Lei Dong and Dechao Bu: Methodology. PeiPei Huo, Zhihao Wang and Wenxin Deng: Software. Jingjia Liu: Visualization. Jin-Cheng Guo: Funding acquisition; Writing - review & editing; Project administration. Yi Zhao: Funding acquisition; Project administration. Yang Wu: Funding acquisition; Writing - review & editing; Supervision.

**Conflict of interest**

The authors have no conflicts of interest to declare.

**Graphical abstract**

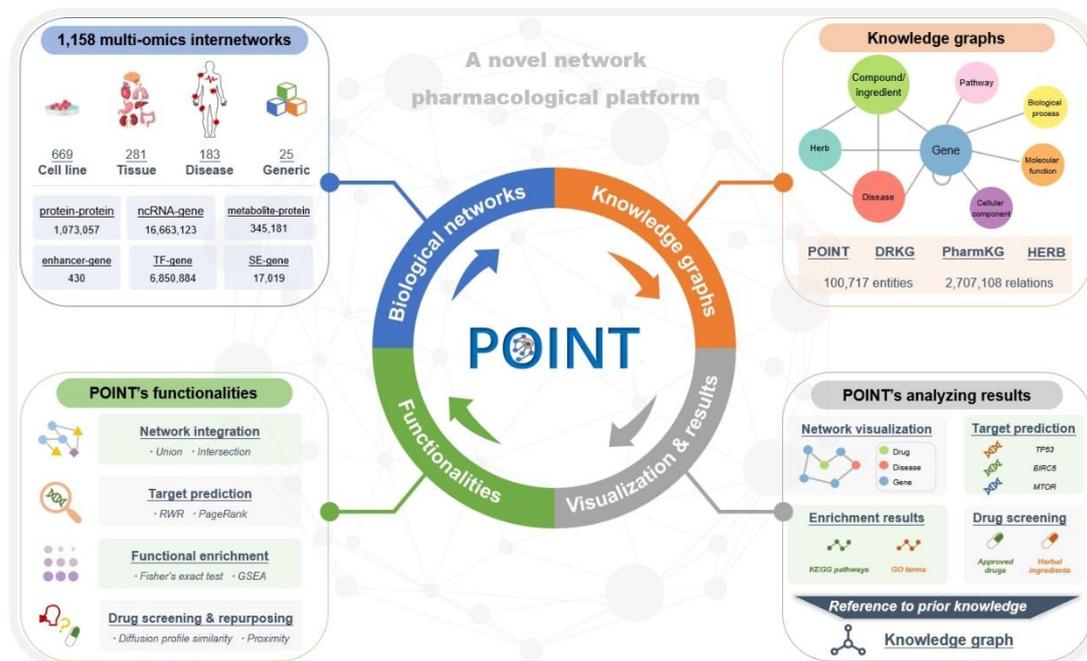

POINT represents a novel network pharmacology platform that synergizes multi-omics biological networks, advanced computational algorithms, and authoritative knowledge graphs (Platform URL: http://point.gene.ac).

Manuscript

Original article

# POINT: a web-based platform for pharmacological investigation enhanced by multi-omics networks and knowledge graphs

**Abstract**

Network pharmacology (NP) explores pharmacological mechanisms through biological networks. Multi-omics data enable multi-layer network construction under diverse conditions, requiring integration into NP analyses. We developed POINT, a novel NP platform enhanced by multi-omics biological networks, advanced algorithms, and knowledge graphs (KGs) featuring network-based and KG-based analytical functions. In the network-based analysis, users can perform NP studies flexibly using 1,158 multi-omics biological networks encompassing proteins, transcription factors, and non-coding RNAs across diverse cell line-, tissue- and disease-specific conditions. Network-based analysis-including random walk with restart (RWR), GSEA, and diffusion profile (DP) similarity algorithms-supports tasks such as target prediction, functional enrichment, and drug screening. We merged networks from experimental sources to generate a pre-integrated multi-layer human network for evaluation. RWR demonstrated superior performance with a 33.1% average ranking improvement over the second-best algorithm, PageRank, in identifying known targets across 2,002 drugs. Additionally, multi-layer networks significantly improve the ability to identify FDA-approved drug–disease pairs compared to the single-layer network. For KG-based analysis, we compiled three high-quality KGs to construct POINT KG, which cross-references over 90% of network-based predictions. We illustrated the platform's capabilities through two case studies. POINT bridges the gap between multi-omics networks and drug discovery; it is freely accessible at http://point.gene.ac/.

## 1. Introduction

Network pharmacology (NP), emerging from the integration of network medicine and polypharmacology, employs biological networks to elucidate pharmacological mechanisms [1,2]. Network medicine utilizes biological network analysis to identify disease-associated genes and potential therapeutic targets, while polypharmacology demonstrates the superior efficacy of multi-target drugs over single-target drugs in treating complex diseases [3,4]. NP integrates the analytical framework of network medicine with the fundamental principles of polypharmacology, establishing itself as a transformative paradigm in modern drug discovery. The NP applied protein-protein interaction (PPI) network has demonstrated significant utility in elucidating therapeutic mechanisms across diverse disease domains, including cardiovascular disorders, oncology, and COVD-19 [5-7].

The increasing discovery of intracellular regulatory elements has expanded the range of potential drug targets beyond proteins to include biomolecules such as transcription factors (TFs) and non-coding RNAs (ncRNAs) [8]. Drug targets include direct targets, which physically bind to biomolecules, and indirect targets, with expression patterns modulated by the drug or its active metabolites. The DrugBank database (v2024) documents 2,227 direct therapeutic targets comprising proteins (2,162), TFs (62), and ncRNAs (3) for small-molecule and biologic agents (**Table S1**) [9]. These findings indicate that PPI networks alone are insufficient for NP analyses and that multi-layer networks should be utilized to integrate biomolecules distributed across various omics layers. For example, Wang *et al.* constructed a multi-layer biological network incorporating proteins, TFs, ncRNAs, and metabolites. Their findings indicated that the direct target of ginsenosides, the TF P53, displays important topological properties throughout the network and participates in various cancer-related signaling pathways [10]. Ma *et al.* developed a multi-omics biological network using proteins and metabolites, which revealed that the TFs Jun and Fos constitute indirect targets in the treatment of leukopenia with Qijiao Shengbai Capsule [11]. Lee *et al.* found that cisplatin-induced acute kidney injury was associated with dysregulation of the indirect targets miR-122 and miR-34a through the ncRNA–mRNA network [12].

In recent years, high-quality biomolecular interaction databases have been published across various omics fields. The HuRI database contains more than 50,000 human protein interactions identified using the yeast two-hybrid technique [13]. GRNdb has compiled human and mouse transcriptome data to construct gene regulatory networks, generating nearly 20 million TF–gene interactions [14]. miRTarBase has aggregated approximately 13,500 experimentally validated ncRNA–gene interactions from published literature [15]. Furthermore, biomolecular interactions have been characterized under diverse conditions, including cell line-, tissue-, and disease-specific contexts, enabling NP studies that more accurately reflect actual pathophysiological states. For example, Wang *et al.* identified donepezil, a candidate drug for Alzheimer's disease, by combining a microglia-specific gene regulatory network with drug–disease association analysis [16]. However, these network layers remain scattered across different sources, lack integration, and have not been adapted for NP analysis.

The growing heterogeneity of biological networks necessitates more robust algorithms for topological analysis while simultaneously addressing the challenge of false-positive predictions. While degree remains the conventional metric for hub node identification in NP, this approach exhibits inherent limitations due to its bias towards densely connected nodes, potentially overlooking biologically significant yet less connected elements [17]. Drugs exert their efficacy in the biological network by binding to direct targets and subsequently influencing indirect targets.

These effects can be characterized as a diffusion process, in which drug-induced perturbations gradually propagate through the network. However, degree alone is insufficient to capture this dynamic feature because it fails to consider the propagation of perturbations via interactions in the network [18]. In order to surmount these limitations, random walk with restart (RWR) has been developed as a network propagation-based method that simulates temporal drug target diffusion dynamics. This method has been demonstrated to be significantly efficacious in disease gene identification and drug screening applications [19]. Confronted with NP prediction results, knowledge graphs (KGs), a powerful tool for storing and retrieving prior knowledge, can undoubtedly refine NP predictions and improve biological interpretability [20]. However, existing large-scale medical KGs lack user-friendly interactive interfaces; successful queries of relevant knowledge often require computational expertise, posing a challenge for NP researchers without sufficient skills. Therefore, a user-friendly platform that integrates NP and KGs is urgently needed.

Here, we developed POINT, a novel platform for NP enhanced by multi-layer networks, advanced algorithms, and KGs (**Fig. 1**). First, we assembled 1,158 multi-omics biological networks covering biomolecules such as proteins, TFs, ncRNAs, enhancers, and metabolites across diverse cell line-specific, tissue-specific, disease-specific, and generic conditions. We then constructed a pre-integrated multi-layer network, including proteins, TFs, and ncRNAs, revealing a significantly improved ability to identify United States Food and Drug Administration (FDA)-approved drug–disease pairs compared to the single-layer network. Second, we implemented advanced algorithms to integrate individual omics-layer networks, predict therapeutic targets, enrich functions for predicted targets, and infer novel drug–disease relationships. RWR outperformed the second-best algorithm, PageRank, achieving 33.1% average improvement in ranking accuracy for the identification of FDA-approved therapeutic targets across 2,002 drugs. Third, we integrated a comprehensive biomedical KG comprising 100,717 entities and 2,707,108 relationships, which conveniently provides a priori knowledge support for NP analysis results. Large-scale evaluation indicated that more than 90% of predicted results could be successfully retrieved from the KG. Finally, we present two use cases to illustrate the capabilities of the POINT platform; our predictions were validated using the KG and published literature.

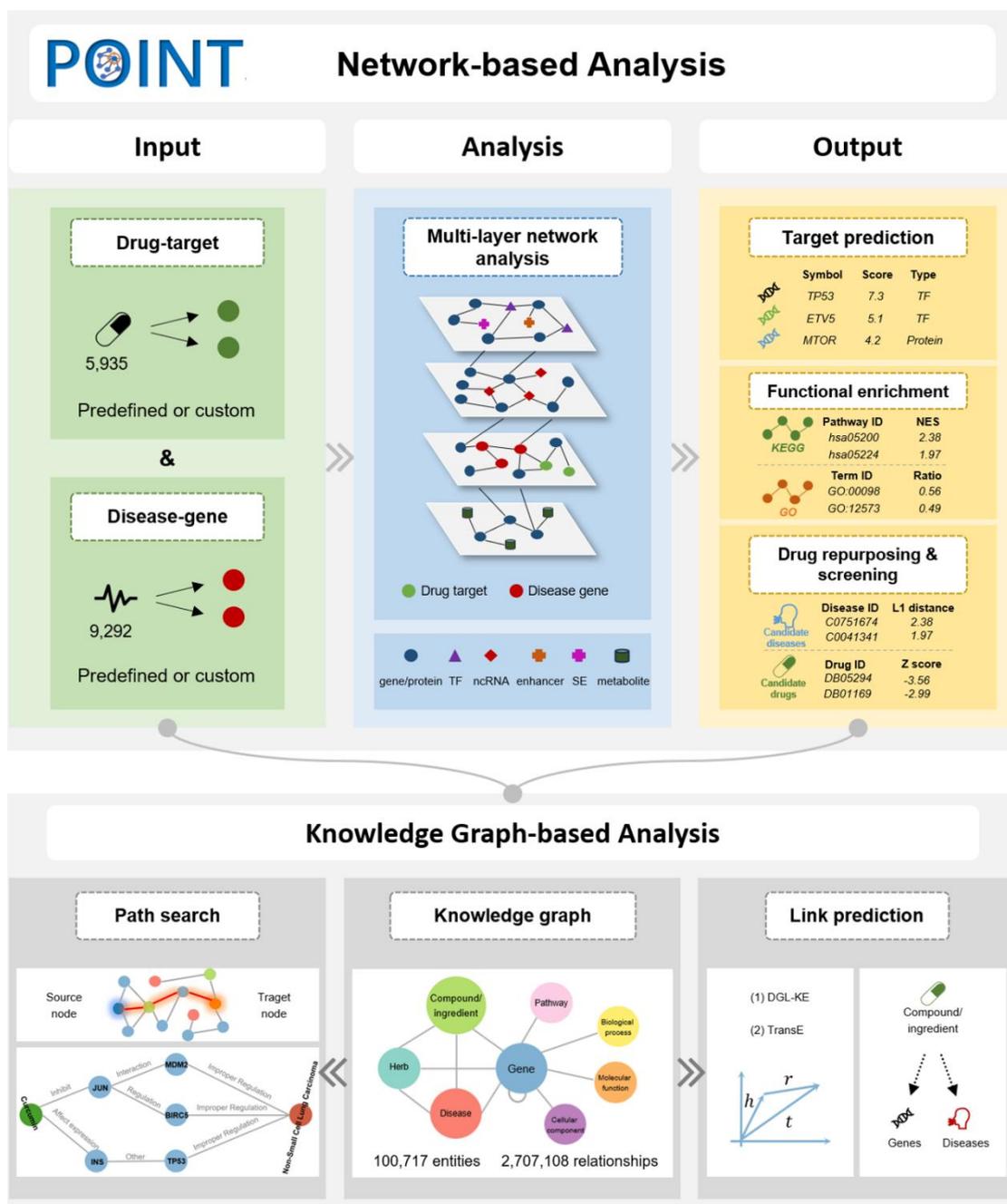

**Figure 1** Flowchart of the POINT platform. The POINT platform comprises two principal analytical modules: network-based analysis and KG-based analysis. Network-based analysis predicts results that can be cross-referenced with a priori knowledge in the KG.

## 2. Materials and methods

*2.1. Data sources*

We initially curated single-layer networks—including PPIs, metabolite–protein interactions, and interactions of critical regulators (*e.g.,* TFs, ncRNAs, enhancers, and super-enhancers) with their targeted genes—from 20 databases (**Table S2**). We performed quality control for each database based on supporting experimental evidence, literature evidence, or computational prediction scores (**Supplementary Fig. S1A**). We retained only human-related networks; our annotations specified cell line, tissue, and disease conditions. Additionally, node identifiers for proteins, TFs, and ncRNAs were standardized to uniform gene IDs using the BioMart tool [21].

We then pre-integrated a multi-layer human network comprising proteins, TFs, and ncRNAs via 13 databases that contained experimentally or literature-validated interactions (**Table S2**). This generic network consists of 18,371 nodes and 388,477 edges (**Supplementary Fig. S1B**). We conducted comparative analyses with single-layer networks to confirm that our network is suitable for general applications. Additionally, computationally predicted multi-layer networks representing specific pathophysiological conditions were collected from the literature [22,23] to accommodate experimental contexts and enhance user convenience. POINT integrates 1,158 biological networks, comprising 1,012 single-layer and 146 multi-layer networks. These networks include 25 generic networks, 669 cell line-specific networks, 281 tissue-specific networks, and 183 disease-specific networks.

To enhance network utilization, we embedded 5,935 drugs, 9,292 diseases, and their associated target genes from the DrugBank [9] and DisGeNET [24] databases. We ensured target relationships by including only expert-validated data. Users can select a drug or disease name from the complete list and use its target genes as seeds to navigate POINT networks. Additionally, users can directly upload drug-related or disease-related genes, allowing flexibility in selecting any gene list of interest.

*2.2. Network-based algorithms and tools in POINT*

POINT provides network-based analytical functionalities: network integration, therapeutic target prediction, functional enrichment of potential targets, and inference of novel drug–disease relationships (*e.g.,* drug repurposing and screening) (**Table 1**).

**Table 1** Algorithms and tools provided in POINT.

| POINT | Functionalities | Algorithms / Tools |
|---|---|---|
| **Network-based analysis** | Network integration | 1) Union; 2) Intersection. |
| | Network visualization (2D/3D) | 1) Echarts; 2) Arena3D. |
| | Target prediction | 1) Random walk with restart (RWR); 2) PageRank; 3) Degree. |
| | Functional enrichment | 1) GSEA; 2) Fisher's exact test; 3) aPEAR. |
| | Drug repurposing & screening | 1) Diffusion profile (DP) similarity; 2) Network proximity. |
| **KG-based analysis** | Path search | 1) NetworkX. |
| | Link prediction | 1) DGL-KE; 2) TransE. |

Users can select one of 146 pre-integrated multi-layer networks or integrate multiple networks from 1,012 single-layer networks for custom applications. When multiple networks are selected, they are first categorized into separate layers, such as the PPI layer or the TF–target gene layer. Networks within each layer are then integrated using either a union or an intersection strategy, as determined by the user. In the union strategy, networks within the same layer are merged to form a single network. In the intersection strategy, interactions occurring at least twice across networks in the same layer are retained, ensuring more stringent integration than the union approach. Finally, integrated networks from all layers are merged into a single network (**Supplementary Fig. S2A**). Because the PPI layer contains extensive information, at least one PPI network is recommended as the backbone for network integration. Additionally, POINT allows users to upload custom networks.

Users can then navigate the integrated network for target prediction using three algorithms: RWR, PageRank, and the degree-based method. These methods prioritize all nodes within the network and identify those with the highest scores as potential targets. (i) The RWR algorithm begins at seed nodes and propagates through the network. With a walk probability of λ (*e.g.,* 0.2), the walker continues to new nodes; with a probability of 1−λ, it restarts at the seed nodes. Seed nodes can represent drug-related or disease-related genes. A smaller λ increases dependence on the seed nodes. After multiple iterations (*e.g.,* 1,000), the algorithm converges and generates a diffusion profile (DP) vector, *r*, which records the frequency of visits to each node (**Supplementary Fig. S2B**). Nodes with the highest frequencies are identified as potential targets. If both drug-related and disease-related genes are provided, RWR runs separately for drug and disease seed nodes. A treatment importance (*TI*) vector is then computed, representing the product of the visiting frequencies in the drug DP $r^{(c)}$ and disease DP $r^{(d)}$ for each node *i,* as shown in **Equation 1**.

$$TI\,(i|c,d) = r_i^{(c)} * r_i^{(d)} \tag{1}$$

(ii) The PageRank algorithm models the limiting distribution of a random walk. It initiates from a randomly selected node, traverses the network over multiple iterations, and assigns each node a PageRank score that reflects its relative importance. (iii) The degree-based algorithm, commonly used in NP analyses and platforms such as CADDIE [25] and Mergeomics [26], ranks nodes according to their degree values, representing the number of connections each node possesses. Nodes with higher degrees in biological networks are considered critical and thus prioritized as potential targets. For the PageRank and degree-based algorithms, users can choose to display only the local ranking of the seed's neighboring nodes. POINT also provides expression levels of these potential targets in normal tissues (Genotype-Tissue Expression Project, GTEx) [27] and cancer tissues (The Cancer Genome Atlas, TCGA) [28].

Next, users can perform Kyoto Encyclopedia of Genes and Genomes (KEGG) [29] and Gene Ontology (GO) [30] functional enrichment analyses for potential targets predicted by POINT using two methods: gene set enrichment analysis (GSEA) [31] and Fisher's exact test [32]. In GSEA, the entire gene list, ranked by target prediction scores, is used as input for functional enrichment analysis. In Fisher's exact test, users select a subset of top-predicted genes (*e.g.,* top 500) for enrichment analysis. Additionally, enriched results are clustered using the aPEAR method [33] which utilizes the Jaccard score to measure similarity between two gene sets. Highly similar gene sets are clustered via Markov clustering, where each cluster contains at least two pathways or terms (**Supplementary Fig. S2C**).

For drug repurposing, users input a drug and a list of candidate diseases. New drugs can be specified according to their related genes. For drug screening, users input a disease and a list of drugs or herbal ingredients. Similarly, new diseases can be defined by their related genes. For each drug–disease pair, POINT maps the associated genes to the pre-integrated multi-layer human network and then measures similarity or distance between the pair using two algorithms (**Supplementary Fig. S2D**). The first algorithm, DP similarity [34], ranks candidate drugs or diseases based on the assumption that similar DPs generated by the RWR algorithm indicate a potential therapeutic relationship. DP similarity is calculated as the L1 distance between the drug DP $r^{(c)}$ and disease DP $r^{(c)}$, as shown in **Equation 2**, where $i$ represents an arbitrary node in the pre-integrated multi-layer human network with N nodes. A smaller DP similarity value (*i.e.,* smaller L1 distance) indicates a greater likelihood of a therapeutic relationship between the drug–disease pair.

$$L1\ distance(c,d) = \sum_{i \in N} \left| r_i^{(c)} - r_i^{(d)} \right| \qquad (2)$$

The second algorithm, network proximity [35], quantifies the distance between gene modules affected by a drug and a disease. Potential therapeutic relationships are inferred when the affected gene modules are in close proximity. Network proximity is calculated based on two gene sets: *C*, representing drug targets, and *D*, representing disease genes. For each gene pair, $g_c$ in *C* and $g_d$ in *D*, $DIS(g_c, g_d)$ denotes the shortest distance between them in the network. Network proximity is determined by averaging all shortest distances, as shown in **Equation 3**.

$$Proximity(C,D) = \frac{1}{\|C\|+\|D\|} \left( \sum_{g_c \in C} min_{g_d \in D} DIS(g_c, g_d) + \sum_{g_d \in D} min_{g_c \in C} DIS(g_c, g_d) \right) \qquad (3)$$

To assess the statistical significance of observed proximity values, a permutation test was performed; Z-scores and P-values were calculated based on the empirical distribution. We generated permutated samples by randomly selecting nodes with the same number and degree values as those in *C* and *D* within the network. Finally, drug–disease pairs with potential therapeutic effects can be sorted based on similarity or distance.

*2.3. Biomedical KGs embedded in POINT*

Additionally, we collected and preprocessed three KGs—DRKG [36], PharmKG [37], and HERB [38]—to construct POINT, a comprehensive KG (**Supplementary Fig. S3**). The POINT KG comprises six distinct entity types: compound/ingredient, disease, gene, herb, pathway, and GO term. These entities collectively represent the critical components of drug action and provide comprehensive support for target prediction, mechanism analysis, and drug screening in NP research. The POINT KG encompasses 100,717 entities and 2,707,108 relationships (**Table S3**). The POINT platform introduced KGs as a priori knowledge to support the interpretation of network-based predictions, including target prediction, drug screening, and drug repurposing. We evaluated POINT KG coverage over network-based prediction results in subsequent analyses.

The POINT platform also provides two KG-based analytical functionalities: path search and link prediction (**Table 1**). Path search allows users to select a KG and define two entities as source and destination nodes, respectively, for identification of paths connecting them within the Python module NetworkX (**Supplementary Fig. S4A**). To optimize computational efficiency, POINT restricts path lengths to a maximum of three entities. If no path is found within this limit, the restriction is lifted. All identified paths are categorized into distinct meta-paths based on entity types and ranked according to the average degree of each meta-path. The average degree of a path is determined by computing the mean degree of all entities in the selected KG. A higher average degree indicates a greater amount of prior knowledge associated with the path.

Link prediction infers new relationships based on existing knowledge within the KG. Users can select a KG and input a query compound/ingredient to predict new associations with genes and diseases (**Supplementary Fig. S4B**). This prediction is performed using the Deep Graph Library Knowledge Embedding (DGL-KE) framework and the TransE algorithm. First, KG triples are divided into a training set (90%), a validation set (5%), and a testing set (5%). The TransE scoring model is then iteratively trained 100,000 times on the training set to optimize parameters in the DGL-KE framework, including embedding dimensions, learning rate, and regularization coefficients; evaluation is performed using the validation set. After training, the weights of all entities and relationships are obtained to infer new associations. We used HITS@10 to evaluate model performance on the testing set; this output represented the proportion of known relationships ranked within the top 10 predicted results. When users predict new relationships between a compound/ingredient and a gene or disease, a new triplet is generated, and a TransE score is assigned. A higher score indicates a greater likelihood that the inferred relationship is valid.

*2.4. Comparison of target prediction algorithms in POINT*

We evaluated the known target recovery performances of RWR, PageRank, and degree-based algorithms embedded in POINT using both single-layer and multi-layer networks. The pre-integrated multi-layer human network (PPI+TF+ncRNA) was regarded as the multi-layer network; its PPI sub-network served as the single-layer network. The evaluation protocol is illustrated in **Fig. 2A**. Step 1: We selected 2,002 drugs (with target numbers ranging from 2 to 20) from the DrugBank database for evaluation. Step 2: We mapped drug targets onto the PPI and PPI+TF+ncRNA networks. Because the RWR algorithm requires specification of seed nodes, we utilized the leave-one-out cross-validation strategy [39]. In each iteration, a known target was temporarily removed from the network as the target to be predicted, and the remaining targets were used as seed nodes. This process was repeated until all targets had been individually removed and re-evaluated as predicted targets. For PageRank and degree-based methods, all known targets were directly mapped onto the network as predicted targets. Step 3: We calculated RWR, PageRank, and degree scores for all predicted targets. Step 4: We evaluated the known target recovery ability of the three algorithms using the Target Retrieve Score (TRS) for each drug, calculated as follows:

$$TRS_{set\ j} = \frac{(\sum Target\ ranking)/Target\ number}{Node\ number_{set\ j}} \tag{4}$$

where *Target ranking* represents the ranking position of predicted targets obtained by different algorithms, *Target number* represents the number of drug targets, and *Node number* defines the ranking range. In Set 1, predicted target ranking was performed across all nodes in the network. In Set 2, predicted target ranking was performed within drug targets and their first- and second-order neighbors. In Set 3, predicted target ranking was performed within drug targets and their first-order neighbors. The purpose of Sets 2 and 3 was to mitigate the advantage of using other known targets as seed nodes in the RWR algorithm.

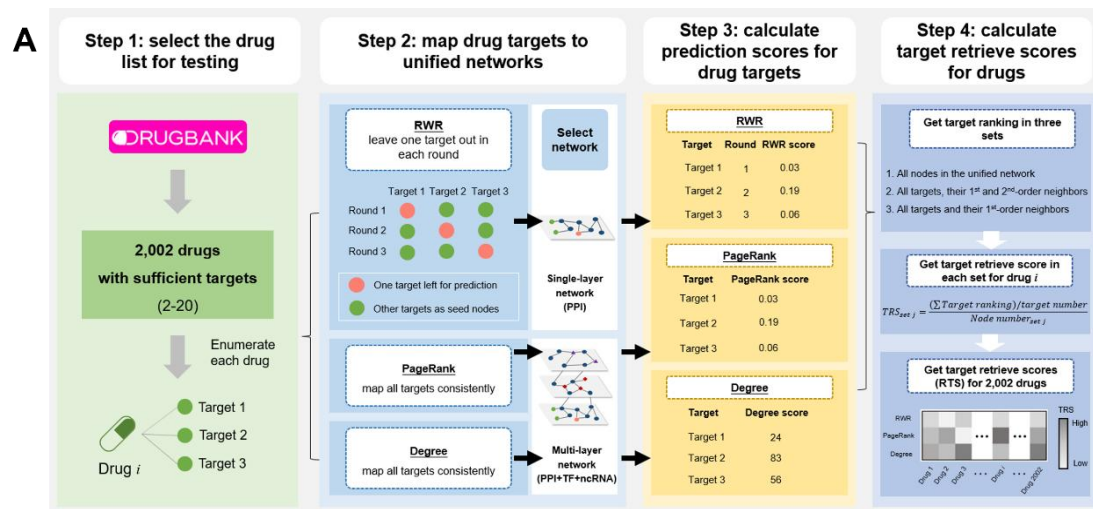

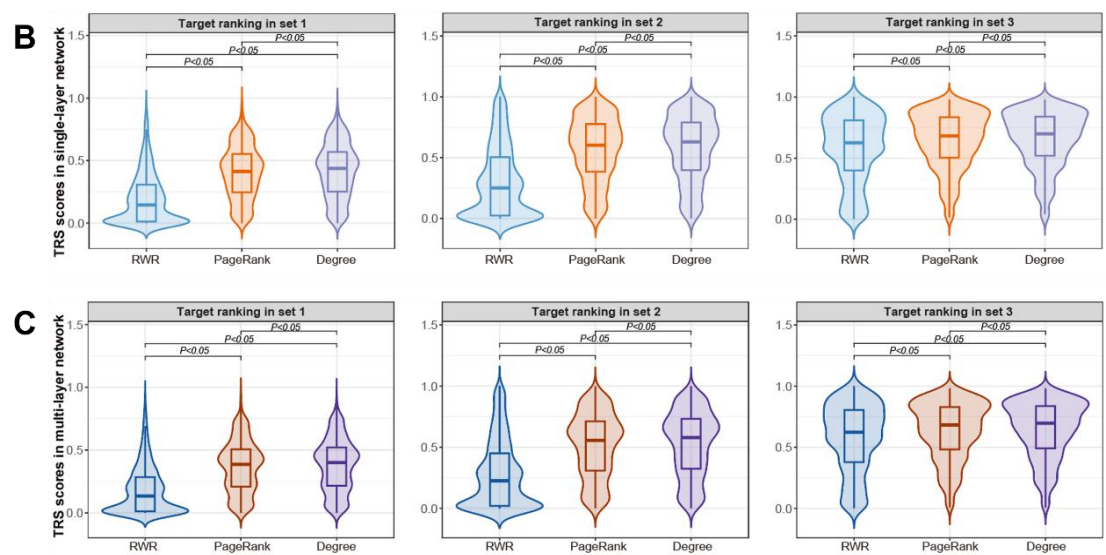

**Figure 2** (A) Flowchart illustrating the comparison of target prediction algorithm performance. A lower average target retrieve score indicates higher prediction accuracy. (B) Target retrieve scores for RWR, PageRank, and degree-based algorithms across different sets in a single-layer network. (C) Target retrieve scores for RWR, PageRank, and degree-based algorithms across different sets in a multi-layer network.

*2.5. Comparison of single and multi-layer networks for drug screening*

To evaluate the performances of single-layer and multi-layer networks in drug screening tasks, we compared the PPI+TF+ncRNA network and its three sub-networks. Three evaluations were conducted, as illustrated in **Fig. 3A**. Evaluation 1 assessed the direct performance of single-layer and multi-layer networks in drug screening. Step 1: We selected four disease types—cancer (non-small cell lung carcinoma, NSCLC), psychiatric disease (Alzheimer's disease), cardiovascular disease (hypertension), and metabolic disease (type 2 diabetes)—along with their corresponding FDA-approved therapeutic drugs. Step 2: For comparison, we selected the PPI, PPI+TF, PPI+ncRNA, and PPI+TF+ncRNA networks. Step 3: We utilized the RWR algorithm to calculate the DPs of drugs and diseases in various networks. To adjust for differences in node numbers across networks, a corrected L1 distance (*i.e.,* L1 distance') was used to calculate DP similarity between drug–disease pairs. Evaluation 2 introduced a random network to eliminate the effect of chance. This random network retained the same numbers of nodes and edges as the PPI+TF+ncRNA network, but edges outside the PPI network were randomly generated. Evaluation 3 incorporated drug–disease pairs without known therapeutic relationships to assess the ability of the PPI+TF+ncRNA network to distinguish between drug–disease pairs with and without therapeutic relationships. Given the computational time cost and scale of testing, we randomly selected 100 drug–disease pairs with known therapeutic relationships from the dataset compiled by Camilo Ruiz [34]. Additionally, we randomly selected 100 drugs and 100 diseases to generate 100 drug–disease pairs without known therapeutic relationships. To mitigate the effect of chance, this process was repeated five times.

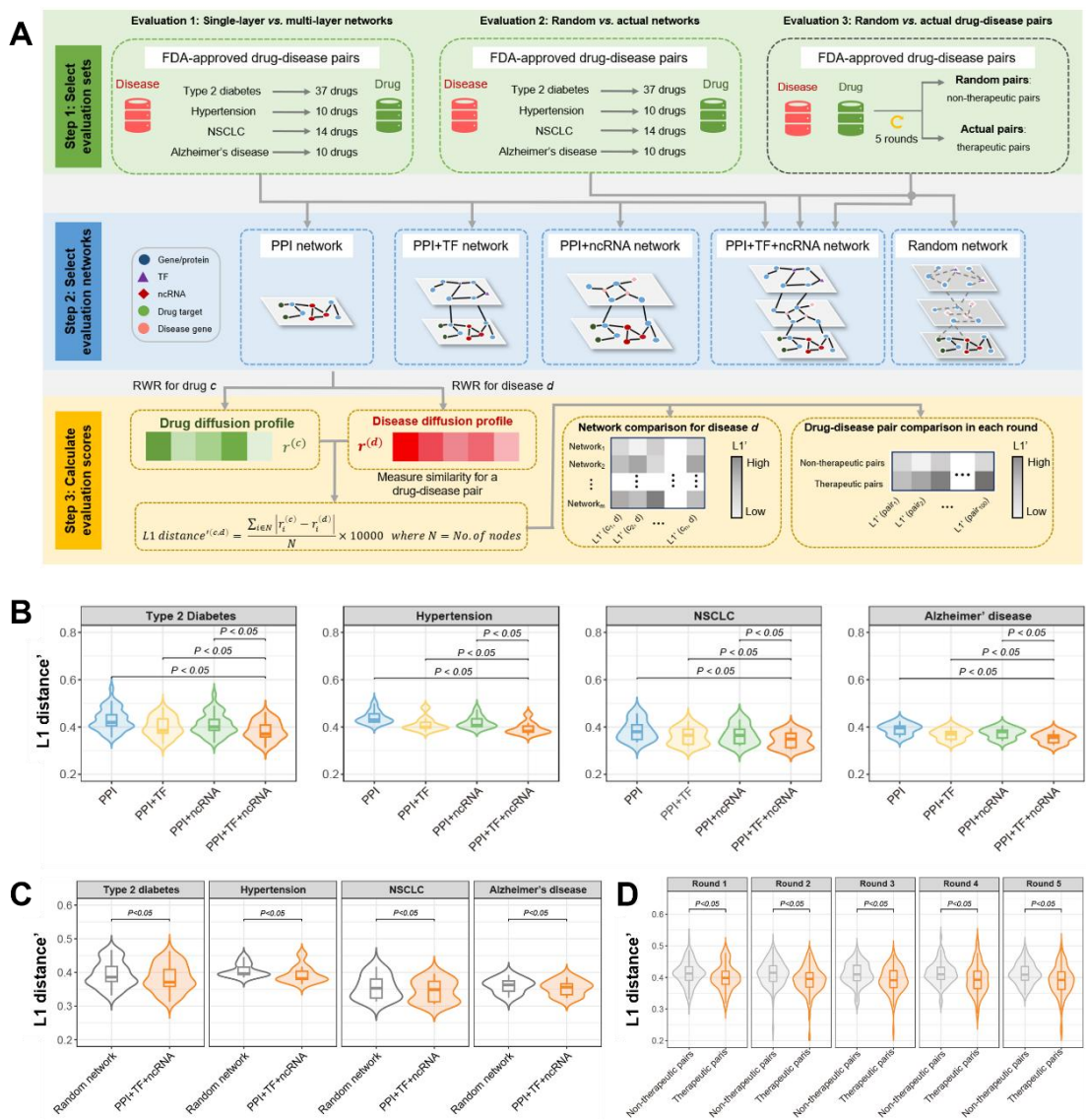

**Figure 3** (A) Evaluation scheme for assessing the performance of the PPI+TF+ncRNA network in predicting drug–disease pairs with therapeutic relationships. A lower L1 distance' suggests a higher likelihood of a therapeutic relationship. (B) Performance comparison of the PPI+TF+ncRNA network

and its sub-networks across different diseases. (C) Comparison of the PPI+TF+ncRNA network with a randomized network containing the same number of nodes but with randomly generated edges. (D) Performance evaluation of the PPI+TF+ncRNA network in identifying drug–disease pairs with therapeutic relationships across five rounds of random sampling.

*2.6. Evaluation of KGs as references for network-based analysis results*

To evaluate the performance of KGs in referencing network-based prediction results, we queried predicted relationships within KGs for target prediction, drug repurposing, and drug screening. The evaluation protocol is illustrated in **Fig. 4A**. Step 1: For target prediction, we applied the RWR algorithm to the 2,002 drugs previously described; the top 100 predicted targets for each drug were retained based on RWR scores. For drug repurposing and drug screening, we used the DP similarity and network proximity algorithms, respectively. In total, 525 diseases and 1,610 drugs were selected from DisGeNET and DrugBank for testing. The DP similarity algorithm retained only the top 100 predicted diseases with L1 distances < 0.8, whereas the network proximity algorithm retained the top 100 predicted diseases with an adjusted $P < 0.05$. Step 2: The results of target prediction, drug repurposing, and drug screening were mapped as drug-predicted target, drug-predicted disease, and disease-predicted drug pairs onto the POINT, DRKG, PharmKG, and HERB KGs. Step 3: The two entities in each predicted pair from the network-based analysis were regarded as source and target entities for path search, with a limit of two intermediate entities. If a path was identified, the predicted pair was considered successfully queried in the KG.

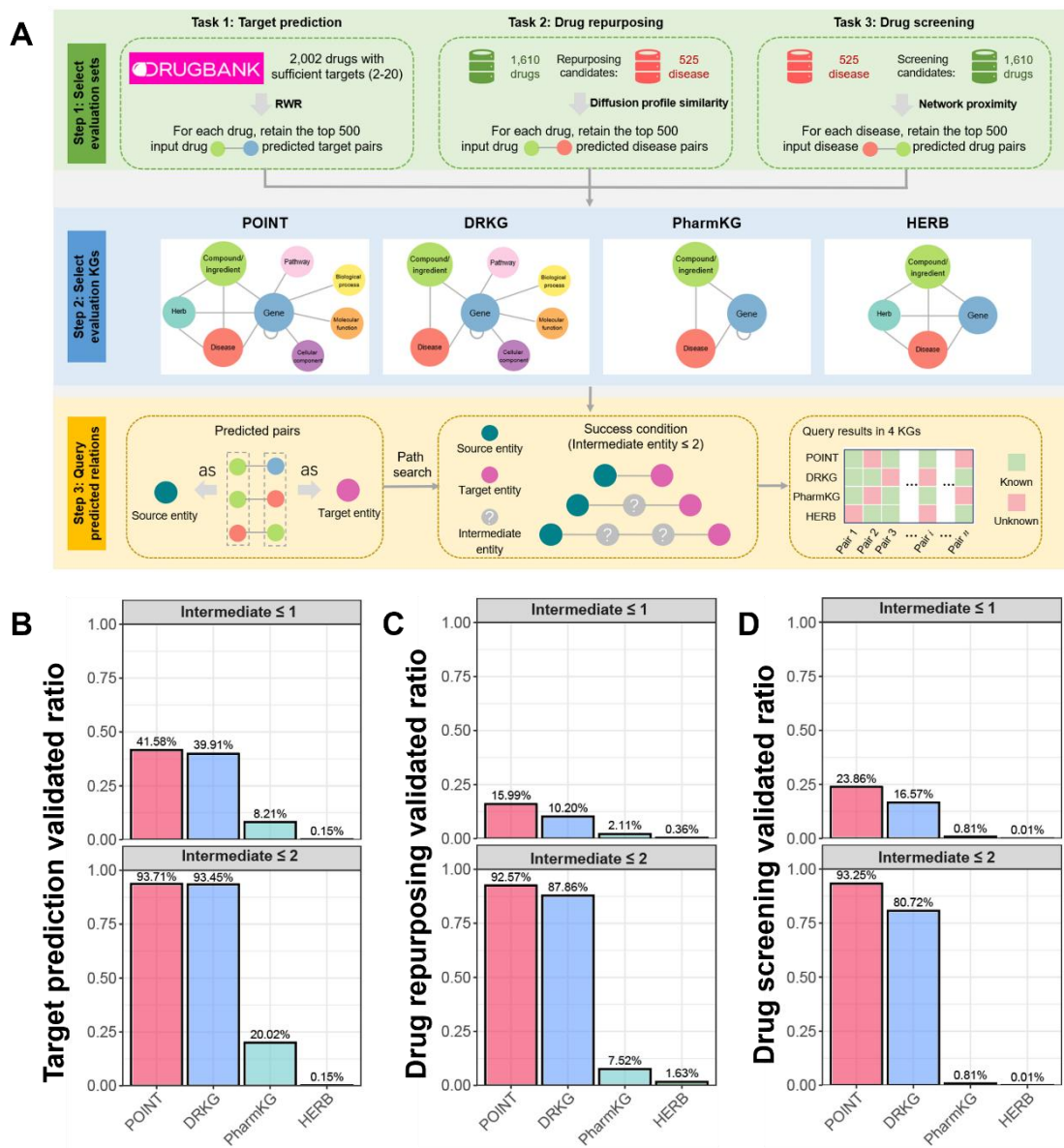

**Figure 4** (A) Flowchart for validating predictions in network-based target prediction, drug repurposing, and drug screening using four KGs. (B) Target prediction validation: proportion of top 100 drug-predicted target pairs for each drug successfully queried in different KGs. (C) Drug repurposing validation: proportion of top 100 drug-predicted disease pairs for each drug successfully queried in different KGs. (D) Drug screening validation: proportion of top 100 disease-predicted drug pairs for each disease successfully queried in different KGs.

## 3. Results

### 3.1. Overview of POINT

The POINT platform consists of two principal analytical modules: network-based analysis and KG-based analysis (**Fig. 1**). The network-based analysis encompasses five core functionalities: (i) Network integration: POINT aggregates thousands of condition-specific multi-omics networks and offers network integration strategies, enabling users to construct a sophisticated multi-layer network for NP research. (ii) Target prediction: By incorporating network topology analysis algorithms, POINT facilitates the identification of potential targets for the input drug within the constructed networks. (iii) Functional enrichment: POINT integrates functional enrichment algorithms to elucidate the effects of drug potential targets on KEGG pathways and GO terms, thereby providing insights into the biological functions modulated by drug interventions. (iv) Drug repurposing: POINT enables the prediction of associations between the input drug and diseases, supporting the discovery of novel therapeutic indications for existing drugs. (v) Drug screening: POINT further predicts drugs or herbal ingredients associated with the input disease, facilitating the identification of promising drug candidates for disease treatment.

In the KG-based analysis, POINT offers users access to four medical KGs: DRKG, PharmKG, HERB, and POINT, with the POINT KG representing a comprehensive integration of the preceding three. Leveraging these KGs, POINT provides two functionalities, including path search and link prediction. The path search feature enables users to query existing paths between entities within the KG, thereby facilitating the identification of established relationships. In contrast, the link prediction capability employs a priori knowledge to infer novel relationships between entities, enabling the extraction of novel insights from prior knowledge. Notably, we have implemented a POINT KG quick query function on the network-based analysis out interface, enabling seamless cross-referencing between prediction results and prior knowledge to refine NP analysis outcomes.

*3.2. The RWR algorithm outperforms PageRank and degree-based algorithms in target prediction*

**Fig. 2B** presents the known target retrieval performances of the RWR, PageRank, and degree-based algorithms for 2,002 drugs in a single-layer network. The TRS metric reflects the ranking position of a drug's known targets within the network; lower values indicate a stronger ability to retrieve known targets. The average TRS values for RWR in sets 1, 2, and 3 were 0.193, 0.303, and 0.580, respectively, which were lower than those for PageRank (set 1=0.403, set 2=0.560, set 3=0.647) and the degree-based algorithm (set 1=0.471, set 2=0.576, set 3=0.656). RWR improved the average ranking position across the three sets by 33.1% and 36.8% compared with PageRank and the degree-based algorithm, respectively. Statistically significant differences in performance between RWR and the other two algorithms were confirmed using the paired Wilcoxon test, reinforcing the superior ability of RWR to recover known drug targets. **Table S4** shows the TRS values for each drug in the single-layer network. For RWR, 1,017, 796, and 245 drug targets ranked within the top 15% in sets 1, 2, and 3, respectively. These numbers exceeded those obtained by PageRank (set 1=302, set 2=204, set 3=65) and the degree-based algorithm (set 1=277, set 2=197, set 3=61). The same analysis was conducted in a multi-layer network, yielding similar conclusions (**Fig. 2C** and **Table S5**). We also examined the effect of λ on TRS, as illustrated in **Supplementary Fig. S5**. Higher λ values allow more extensive network exploration but do not improve the recovery of known targets. Across all three datasets, optimal performance was achieved when λ = 0.2. Notably, all three algorithms achieved lower TRS values in the multi-layer network than in the single-layer network, indicating the former's superiority in terms of identifying drug targets.

*3.3. The multi-layer network outperforms the single-layer network in drug screening*

We used L1 distance' to measure DP similarities between four diseases and their corresponding therapeutic drugs in both the single-layer network (PPI) and three multi-layer networks (PPI+TF, PPI+ncRNA, PPI+TF+ncRNA) (**Fig. 3B**). A smaller L1 distance' indicates more remarkable DP similarity, reflecting a higher probability that the drug and disease induce analogous perturbations within the network, suggesting an increased likelihood of a therapeutic relationship between the drug and the disease. We found that the DP similarity of the multi-layer networks was consistently higher than that of the single-layer networks (**Table S6**). Among multi-layer networks, PPI+TF+ncRNA outperformed both PPI+TF and PPI+ncRNA, suggesting that inclusion of enriched regulatory elements enhances PPI network performance in drug screening. To further assess this improvement, we compared DP similarity between the random network and PPI+TF+ncRNA. The latter exhibited superior performance (**Fig. 3C**), demonstrating that the observed enhancement resulting from integration of TF- and ncRNA-related regulatory elements into the PPI network was not due to chance. Additionally, we evaluated the ability of the PPI+TF+ncRNA network to distinguish between drug–disease pairs with and without therapeutic relationships (**Fig. 3D**). The PPI+TF+ncRNA network exhibited significantly higher DP similarity scores for drug–disease pairs with known therapeutic relationships across all five random sampling tests (**Table S7**). All observed advantages were statistically significant, as confirmed by the Wilcoxon test.

*3.4. The POINT KG provides a priori knowledge to support network-based analysis results*

We generated an extensive set of predictive relationships using network-based algorithms to evaluate the reference utility of the POINT KG in refining and validating the prediction outcomes. The top 100 results from the target prediction, drug repurposing, and drug screening tasks yielded 200,200 drug–predicted gene pairs (**Table S8**); 136,726 drug–predicted disease pairs (**Table S9**); and 52,418 disease–predicted drug pairs (**Table S10**). The query success rates for these predicted pairs in POINT, DRKG, PharmKG, and HERB are shown in **Fig. 4B–D**. As the number of intermediate entities increased, the query success rate for all KGs tended to improve. When the number of intermediate entities was limited to two, POINT KG achieved an average query success rate of 93.2% across all three tasks, outperforming the other KGs. To validate these findings, we queried the top 500 predicted pairs from each task using the same approach; these additional queries yielded similar results (**Supplementary Fig. S6**). Our results indicate that POINT KG provides broader medical knowledge coverage compared with other KGs. However, a small proportion of predicted relationships remain unconfirmed; researchers should interpret the results cautiously.

*3.5. POINT's user-friendly interface*

*3.5.1. Network-based analysis*

The POINT platform features a user-friendly interface for network-based analysis, KG-based path search, and link prediction. Network-based analysis involves three steps (**Fig. 5A**): (1) Input drugs and diseases, select from 5,935 drugs and 9,292 diseases curated in POINT, or custom upload drug and disease-related genes; (2) Choose the underlying biological network via pre-integrated networks, multiple network integration (union or intersection), or custom network upload; and (3) Set the analysis tasks, algorithms, and parameters. Computational time typically ranges from 2 to 8 minutes, depending on network size and task complexity.

The network-based analysis output interface, when all tasks are selected, comprises the following sections (**Fig. 5B**): (i) Network overview, displaying second-order input drug neighbors and first-order input disease neighbors within the constructed biological network, with 2D and 3D visualization options; (ii) Target prediction, listing input drug-related predicted targets; (iii) Functional enrichment analysis, highlighting significantly enriched KEGG pathways and GO terms associated with predicted targets, with representative pathways selected using the aPEAR algorithm [33] to streamline analysis; (iv) Drug repurposing, identifying potential diseases treatable by the input drug, visualized as a word cloud of the top 50 disease classes; and (v) Drug screening, identifying potential therapeutic drugs for the input disease, summarized as a word cloud of the top 50 drug classes. All predicted relationships can be cross-referenced with prior knowledge via POINT KG.

*3.5.2 KG-based path search analysis*

The path search function identifies connections between two entities within the KG and involves three steps (**Supplementary Fig. S7A**): (1) Choose the KG; (2) Choose the source node's entity type and name; and (3) Choose the target node's entity type and name. Computational time typically ranges from 2 to 12 minutes, depending on the KG selected.

The output interface includes meta-path classification, path search results, and KG visualization (**Fig. 5C**). Meta-path classification summarizes paths sharing common node and relationship types, with intermediate node and relationship types varying across meta-paths. Individual paths within a meta-path are displayed in the path search results panel, with the average degree of the path and higher average degrees indicating more sufficient prior knowledge. The user can select multiple paths to visualize.

*3.5.3. KG-based link prediction analysis*

The link prediction function assesses the likelihood of a compound/ingredient forming new connections with genes and diseases in the KG. The analysis involves three steps (**Supplementary Fig. S7B**): (1) Choose the KG, (2) Choose a compound/ingredient, and (3) Choose the relationship type for predicting new links. Computational time typically ranges from 10 to 15 minutes, depending on the KG.

The output interface includes prior knowledge visualization, displaying existing KG connections for the input compound/ingredient, and novel knowledge prediction, presenting predicted relationships between the compound/ingredient and other gene or disease entities (**Fig. 5D**).

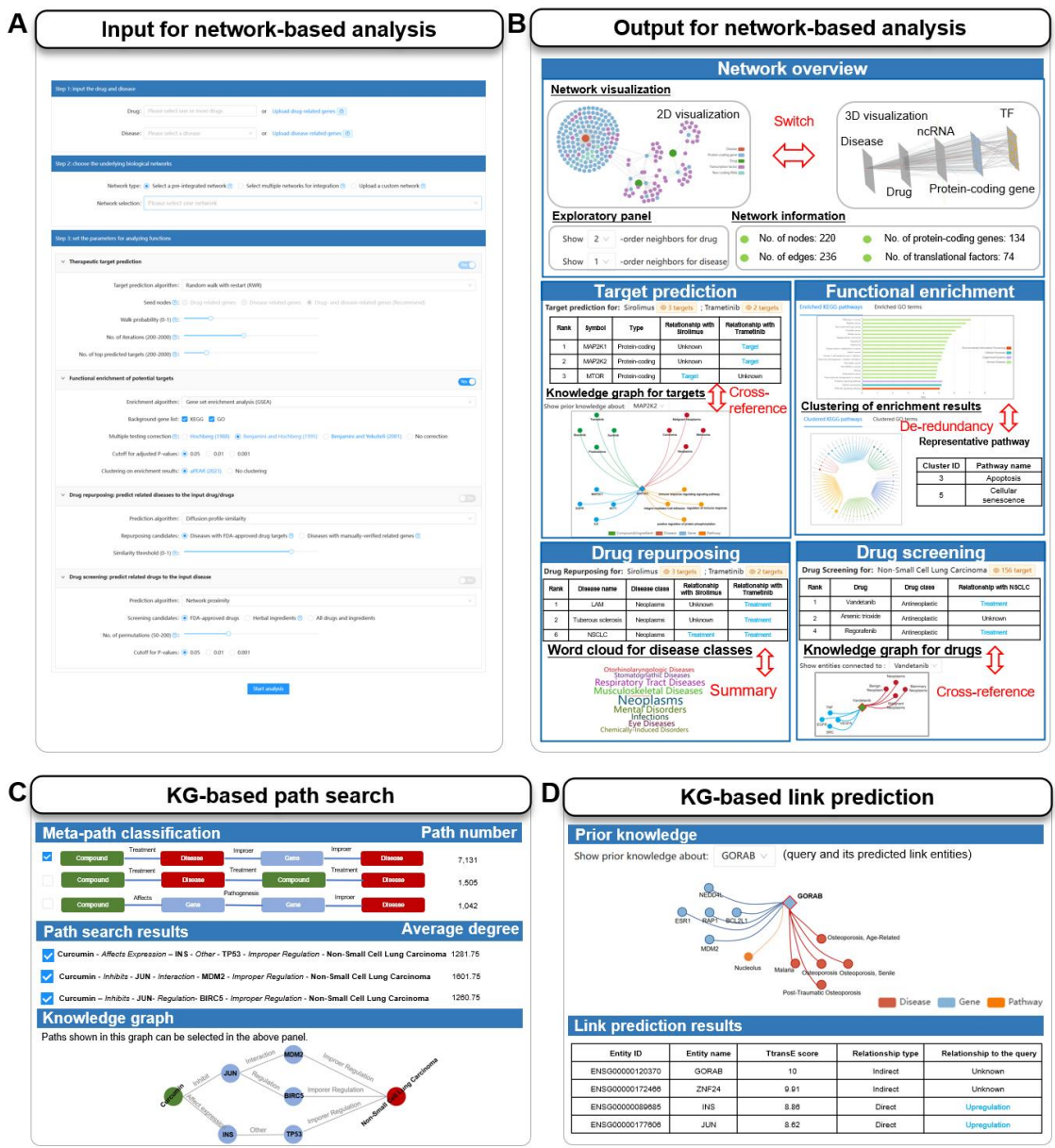

**Figure 5** Input and output of the POINT platform. (A) Input interface for network-based analysis. (B) Output interface for network-based analysis. (C) Output interface for KG-based path search. (D) Output interface for KG-based link prediction.

*3.6. Use Cases*

*3.6.1. Use 1: Exploring dual drug combination therapy for type 2 diabetes using a pre-integrated multi-layer human network*

The combination of metformin and semaglutide represents an effective therapeutic strategy for type 2 diabetes [40]. This case study illustrates a pre-integrated multi-layer human network application for network-based analysis, complemented by KG-based analysis to validate network-derived predictions. Detailed parameters can be accessed via the input interface's 'Load pre-integrated network demo' option.

**Fig. 6A** shows the core sub-network of metformin, semaglutide, and type 2 diabetes within the pre-integrated multi-layer human network. **Fig. 6B** displays the top five predicted targets; GLP1R ranks highest as a known semaglutide target. GPD1 and PRKAB1, ranked fourth and fifth, respectively, are known metformin targets. FAS, ranked second, has no direct association with either drug; however, a KG query identified FAS as a key enzyme in fatty acid synthesis and a major component of the insulin regulatory pathway. GCG, ranked third, is not directly linked to the drugs but plays a critical role in blood glucose regulation. Its tissue-specific expression reduces the risk of off-target effects, making GCG an ideal target for glycemic control in type 2 diabetes.

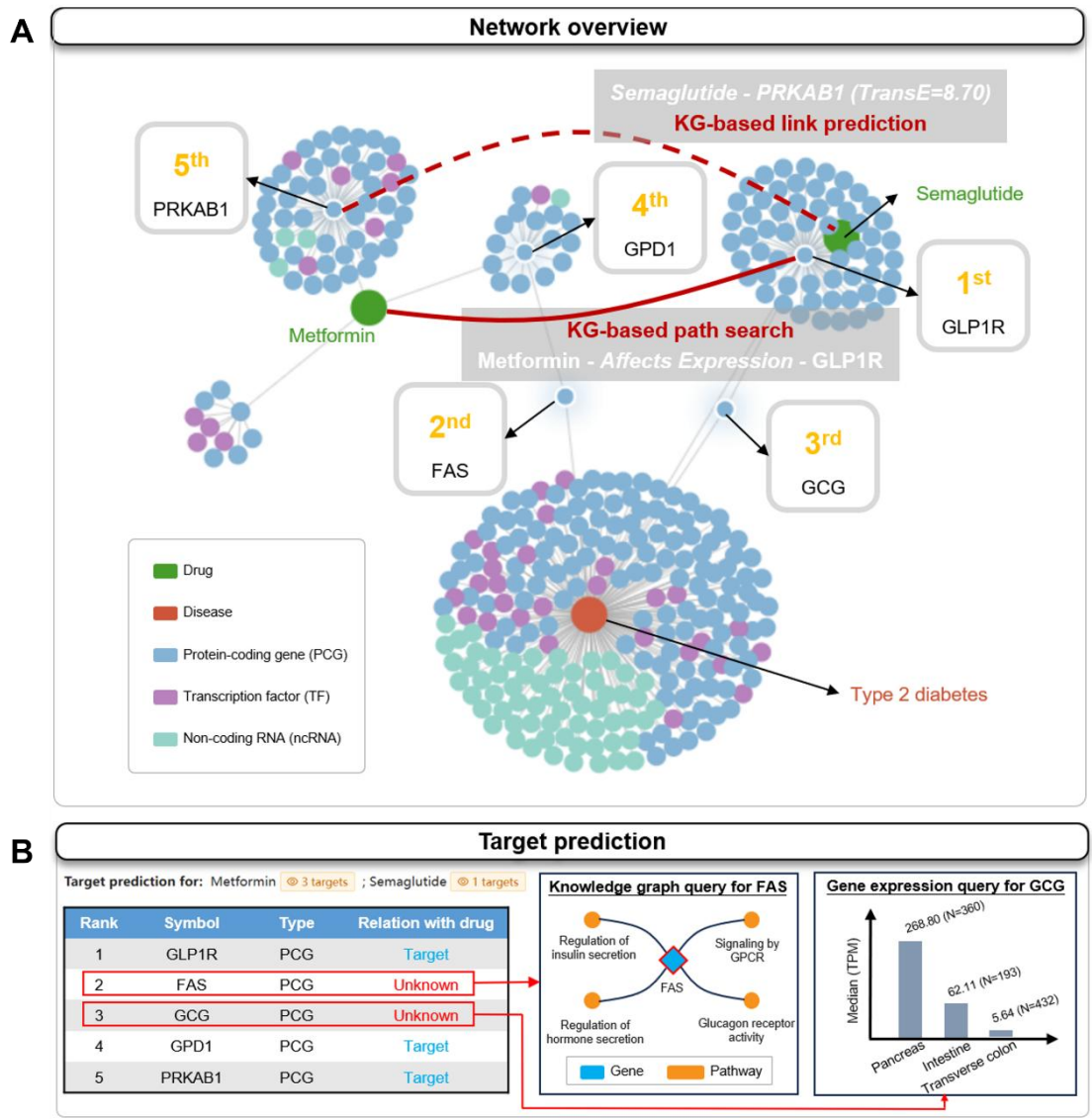

**Figure 6** Exploring the mechanism of metformin and semaglutide combination therapy for type 2 diabetes using network-based and KG-based analyses. (A) Core sub-network of metformin, semaglutide, and type 2 diabetes within a pre-integrated multi-layer human network. The red solid line represents a result from KG-based path search. The red dashed line represents a result from KG-based link prediction. (B) Analysis of the top five potential targets identified through target prediction, validated using KG query and gene expression data.

The KG-based path search results (**Fig. 6A**) demonstrate that semaglutide modulates GLP1R expression to exert its therapeutic effect on type 2 diabetes, aligning with network-based analysis. Metformin also influences GLP1R expression, supported by evidence of its indirect activation through increased endogenous GLP-1 secretion, enhanced GLP-1 signaling, and suppressed hepatic gluconeogenesis, contributing to glucose reduction [41]. KG-based link prediction (**Fig. 6A**) suggests a potential connection between semaglutide and PRKAB1, a direct metformin target, with evidence indicating semaglutide indirectly modulates PRKAB1 via GLP-1 receptor activation, enhancing insulin sensitivity and reducing blood glucose [42]. These findings complement network-based analysis, offering mechanistic insights into the synergistic hypoglycemic effects of the dual-drug combination.

*3.6.2. Use 2: Exploring dual drug combination therapy for NSCLC using a custom-integrated multi-layer tissue-specific network*

The target prediction accuracy of the RWR algorithm can be improved using tissue-specific networks, as outlined in **Supplementary Fig. S8**. POINT integrates condition-specific networks and provides advanced network integration strategies. This case study exemplifies the construction of a multi-layer tissue-specific network for a specific research context and explores the mechanistic basis of sirolimus and trametinib combination therapy in NSCLC [45]. Detailed input parameters are accessible via the 'Load custom network demo' option on the input interface.

**Fig. 7A** presents the core sub-network of the multi-layer lung-specific network. **Fig. 7B** displays the target prediction results; known targets of sirolimus and trametinib—MAP2K1, MAP2K2, and mTOR—were ranked in the top three. The gene expression query function indicated that SMARCA4, ranked fourth, was aberrantly overexpressed in both lung adenocarcinoma and lung squamous cell carcinoma. A recent study demonstrated that SMARCA4 plays critical roles in NSCLC development, metastasis, and drug resistance, making it a novel therapeutic target [43]. E2F6, ranked fifth, was identified via the KG query function as a regulatory component of cell cycle progression. Studies have shown that E2F6, a member of the E2F family of TFs, is highly expressed in various cancers where it affects cell cycle progression and mitosis; thus, it may have potential as a therapeutic target [44].

The non-small cell lung cancer pathway (hsa05223) ranked third, highlighting the potential efficacy of the dual-drug combination in NSCLC (**Fig. 7C**). Enrichment results also included the mTOR (hsa04150) and MAPK (hsa04010) signaling pathways, where sirolimus and trametinib targets function as primary regulators. The PI3K-Akt signaling pathway (hsa04151) plays a critical role in dual-agent synergy by mitigating sirolimus resistance [45]. Intriguingly, the longevity-regulating pathway (hsa04211) was also enriched, consistent with evidence that sirolimus and trametinib extend lifespan in *Drosophila* [46].

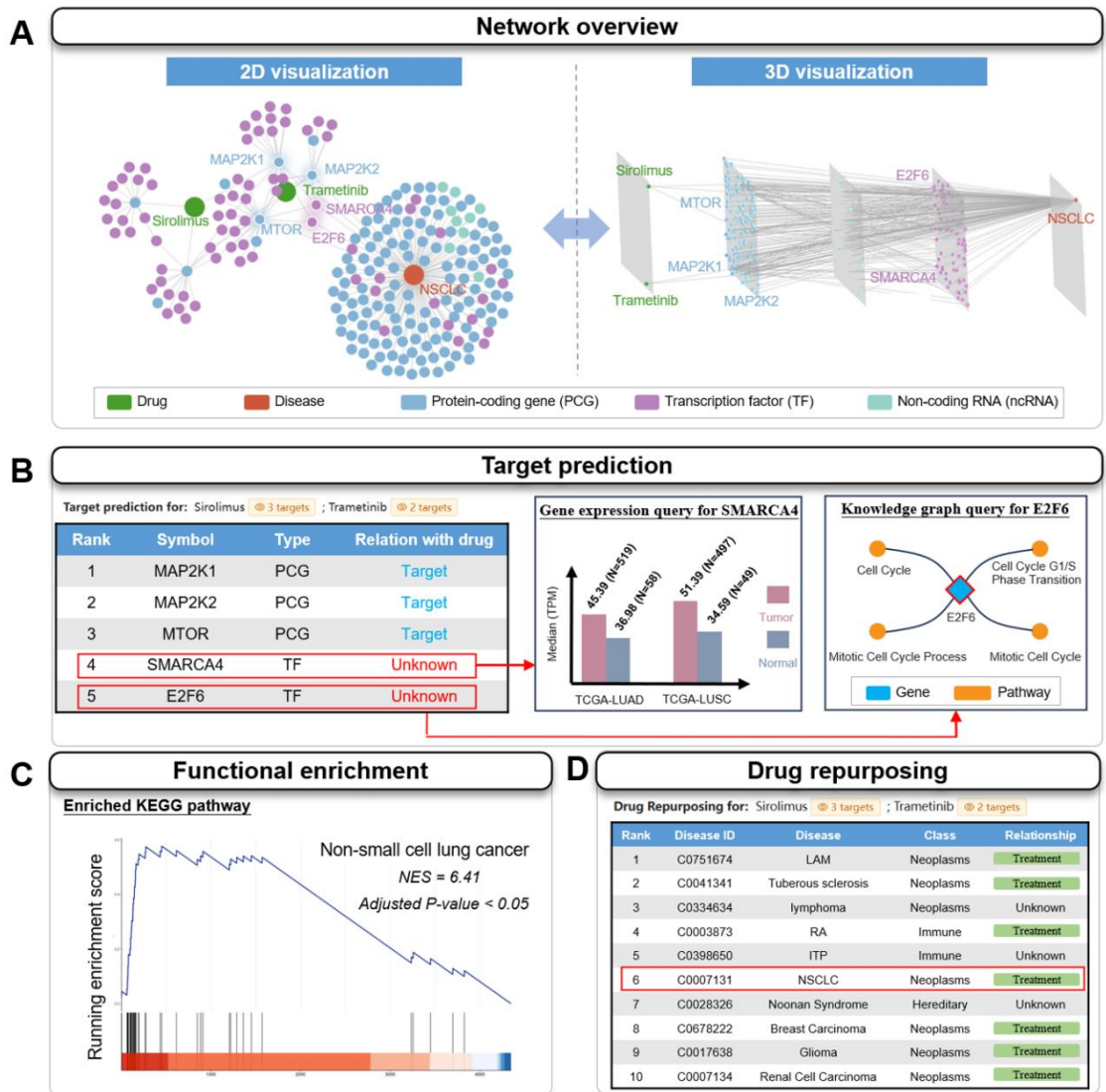

**Figure 7** Exploring the mechanism of sirolimus and trametinib combination therapy for NSCLC using network-based analysis. (A) Core sub-network of sirolimus, trametinib, and NSCLC within a multi-layer lung-specific network. (B) Analysis of the top five potential targets identified through target prediction, validated using KG query and gene expression data. (C) Functional enrichment results based on GSEA. (D) Drug repurposing results, the relationship column represents the prior knowledge of input drugs and predicted diseases in POINT KG.

The drug repurposing results (**Fig. 7D**) revealed that seven of the top 10 predicted treatable diseases for the sirolimus–trametinib combination were tumor-related, including NSCLC, ranked sixth. Notably, seven of these diseases exhibited direct therapeutic relationships with either sirolimus or trametinib in the POINT KG.

## 4. Discussion

The increasing availability of cell line-, tissue-, and disease-specific omics data has provided unique insights into biological systems; however, each omics layer offers only a partial perspective, often insufficient for fully elucidating the complex mechanisms underlying drug treatment effects. This underscores the necessity of integrating multi-omics data to construct multi-layered biological networks, thereby advancing NP research. Existing network-based analysis platforms, such as CADDIE [25], Mergeomics [26], and SmartGraph [47] (**Table S11**), do not systematically integrate these resources. These platforms predominantly provide classical network analysis tools, such as degree-based analysis and shortest path algorithms. Due to the incompleteness of biological networks, these local network-based approaches tend to prioritize highly connected nodes, thus underestimating the importance of loosely connected nodes in biological systems. In contrast, information flow simulation methods (*e.g.,* RWR based on network propagation) leverage the entire network topology, partially addressing these limitations [17]. POINT integrates these resources and algorithms for the first time and provides a user-friendly interface that is exceptionally convenient for NP researchers without a computer background.

Multiple rigorous quality control procedures were implemented to ensure the reliability of the collected biological networks. Additionally, a pre-integrated multi-layer human network was constructed, and various test tasks were conducted within this network to assess its performance. We compared target identification performances across RWR, PageRank, and degree-based algorithms. The results demonstrated that RWR exhibited superior potential for identifying drug targets compared with PageRank and degree-based algorithms. The PageRank and degree-based algorithms are classical network topology analysis methods used to extract key nodes in scale-free networks, which are essential for maintaining network stability due to their high connectivity and topological importance. However, drug targets do not necessarily correspond to nodes with high topological significance. RWR considers global network topology by assigning specific weights to multiple starting nodes, thus preventing the over-ranking of highly connected nodes that are not drug targets. Furthermore, ablation experiments were conducted using the pre-integrated multi-layer human network to predict whether drug–disease pairs exhibited a therapeutic relationship. The results indicated that the incorporation of TFs and ncRNAs into the PPI network enhanced the ability of the DP similarity algorithm to identify corresponding FDA-approved drugs in multiple disease types. Topologically, the RWR-based DP approach effectively simulates drug- and disease-induced network perturbations, highlighting that the integration of multi-layer biological networks with advanced topological algorithms significantly improves the accuracy and reliability of drug screening.

As a pioneering NP platform, POINT resolves three fundamental NP limitations: (1) multi-omics network integration surpassing single-proteome PPI constraints; (2) RWR implementation mitigating degree-based prediction biases; (3) medical KG construction facilitating NP prediction validation and mechanistic interpretation. POINT will undergo regular maintenance and updates to incorporate the latest biological network resources. The ongoing enhancement of POINT is expected to substantially improve the efficiency and reproducibility of

NP research.

## Availability

The POINT platform is freely available at http://point.gene.ac or http://bioinfo.org/point. The biological networks used in this study are available at http://point.gene.ac/Download/.

Figure

Click here to access/download;Figure;POINT_figures.pptx

**Figure 1**

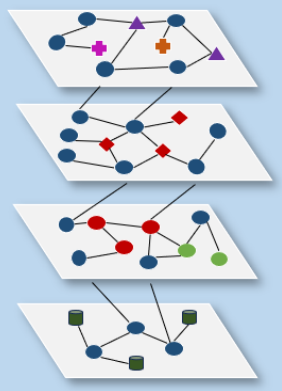

Figure 2

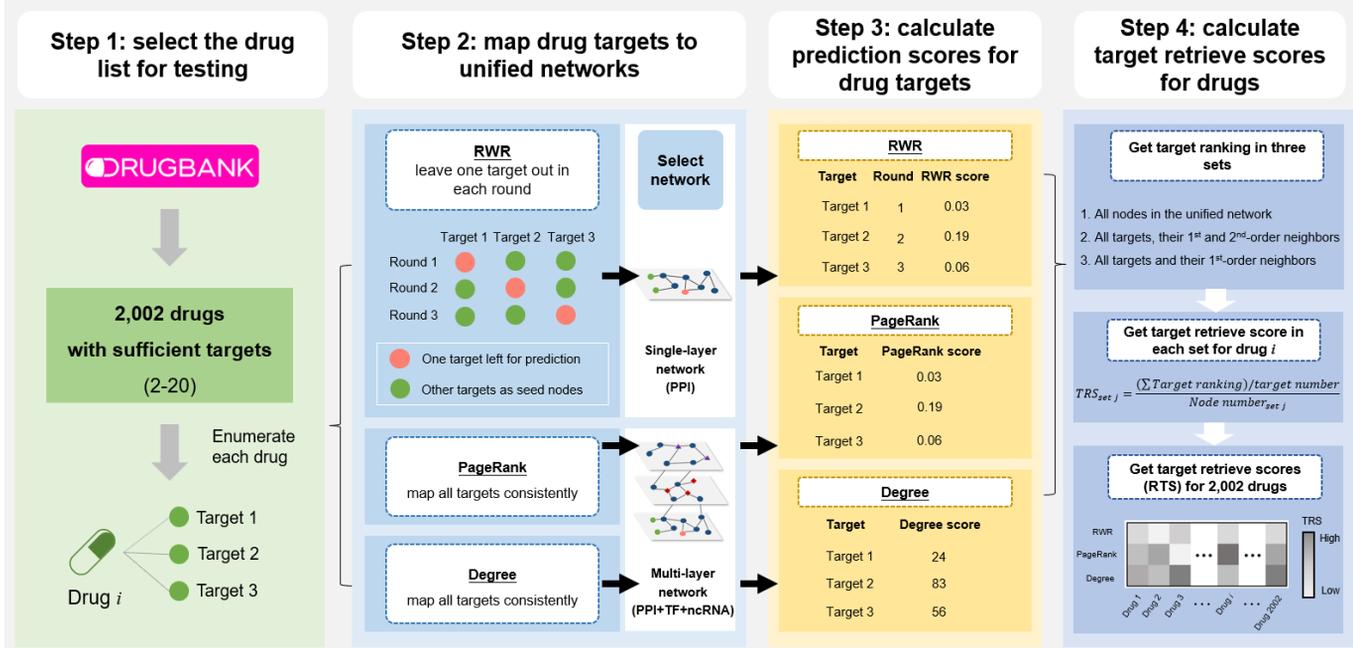
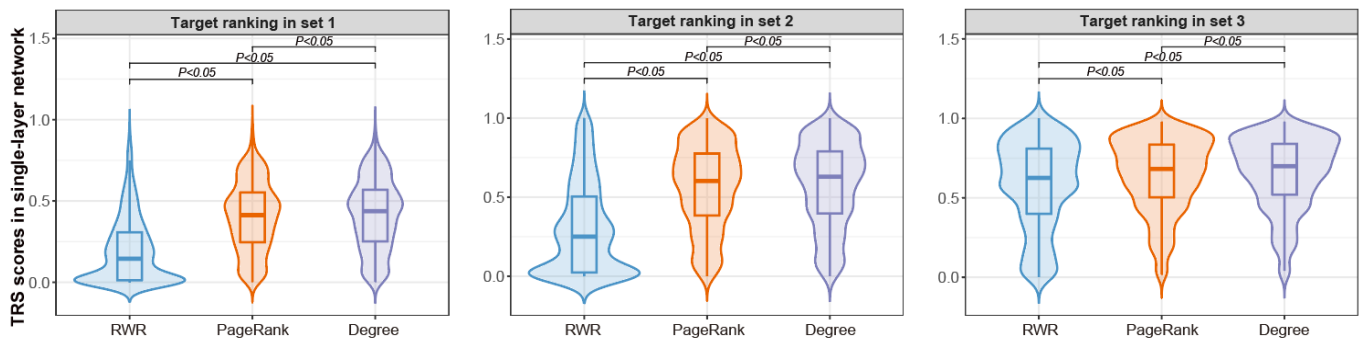
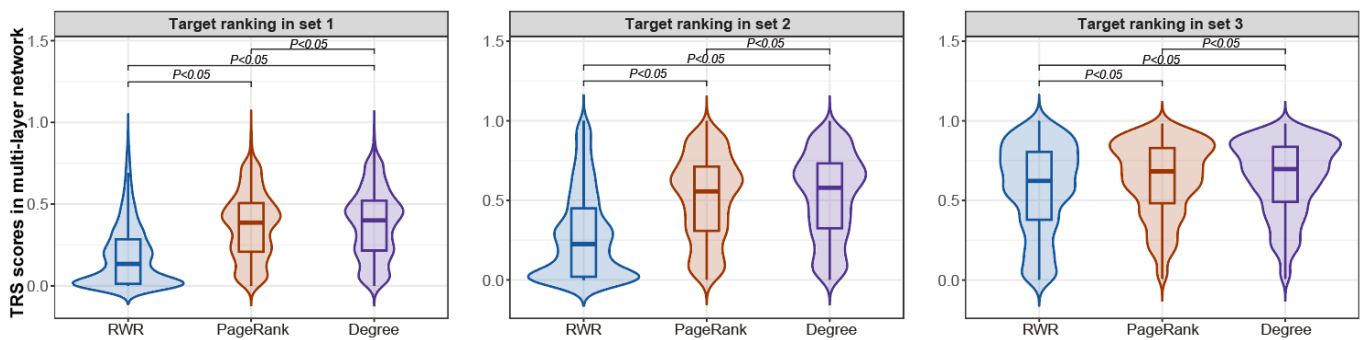

**Figure 3**

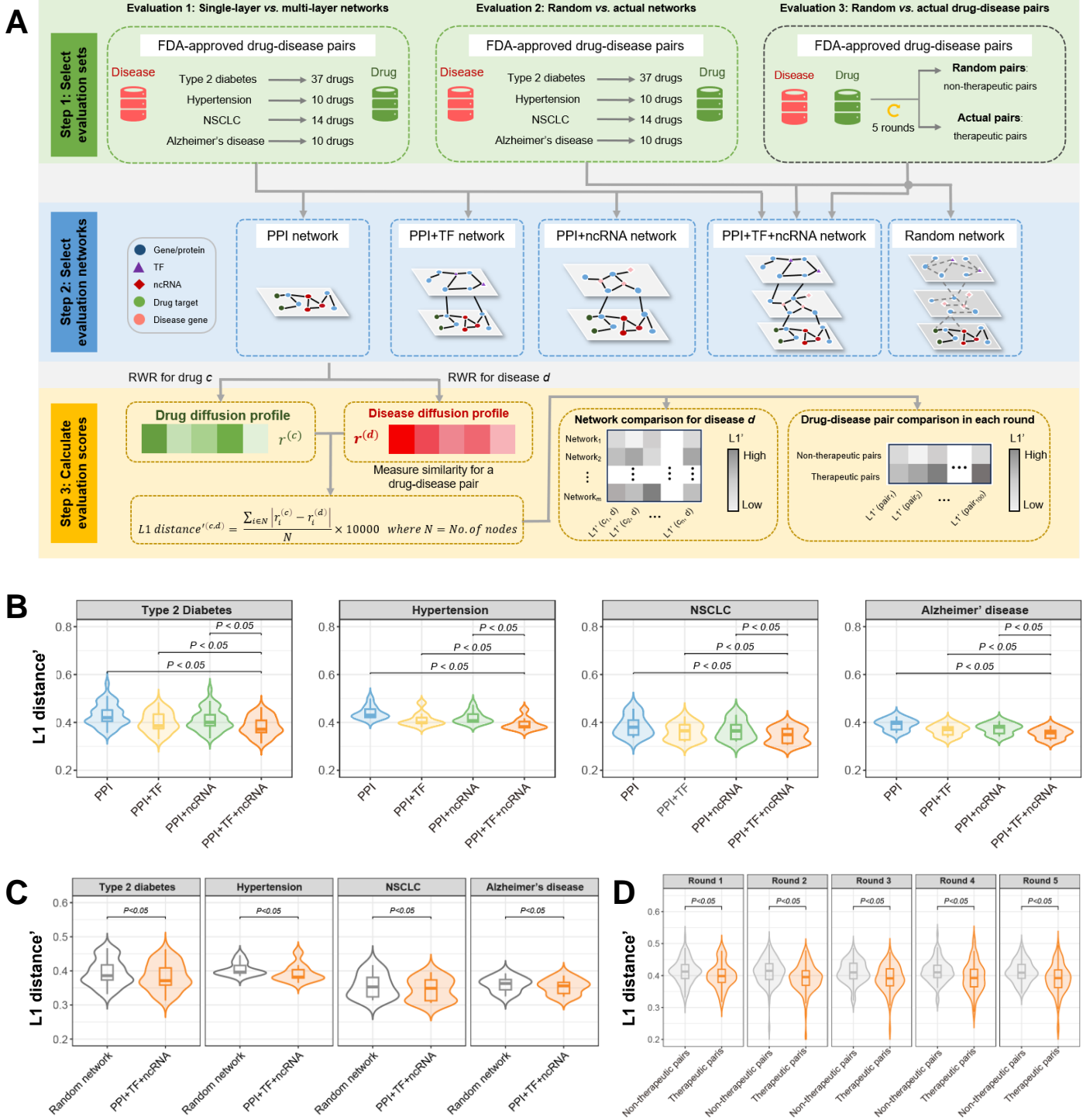

Figure 4

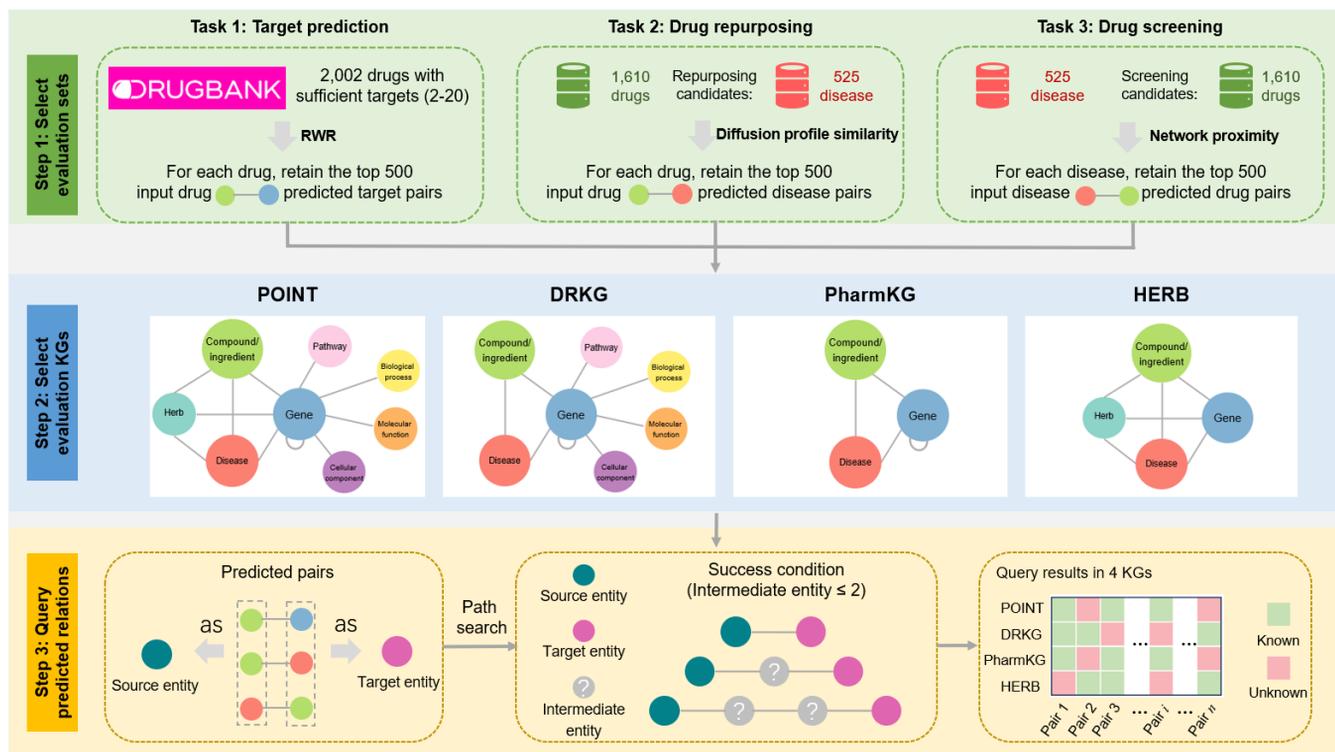

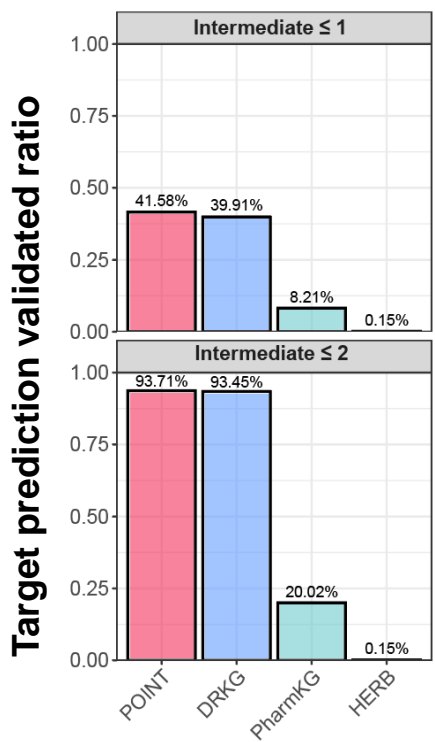
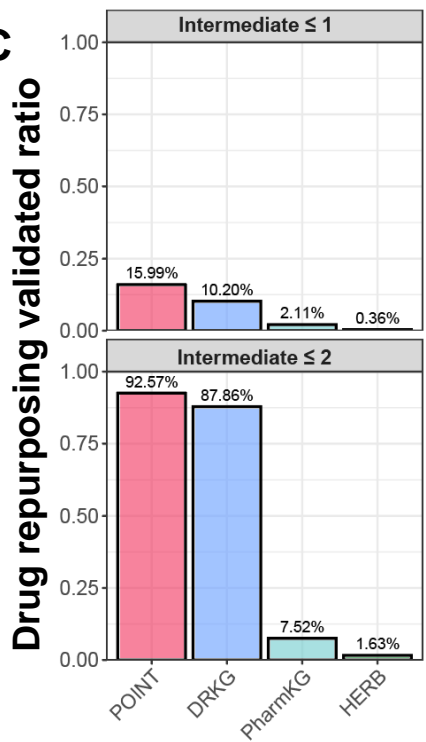
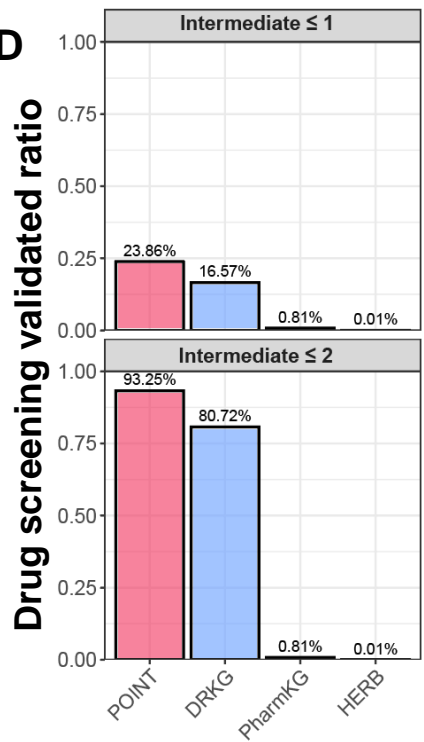

**Figure 5**

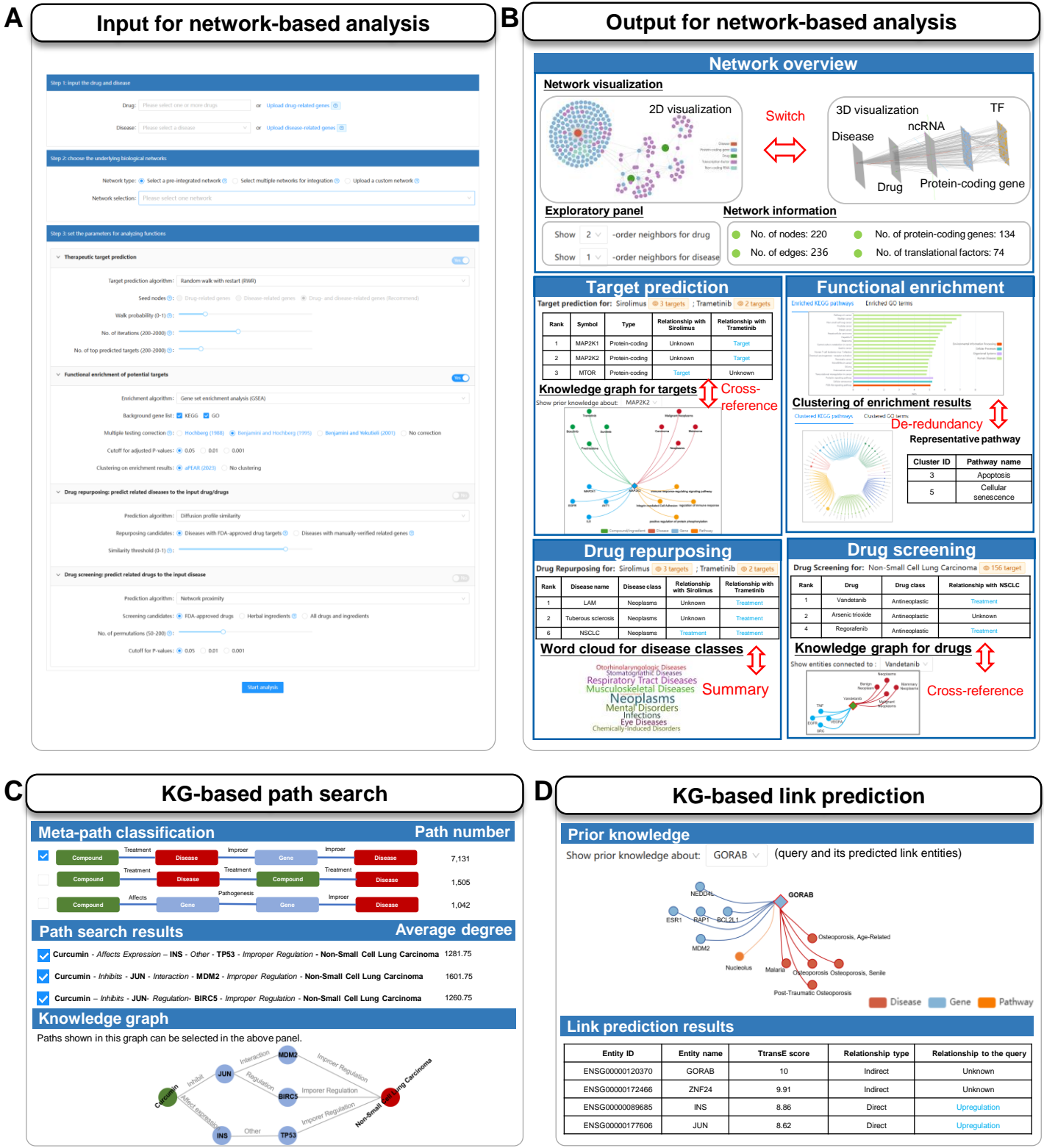

**Figure 6**

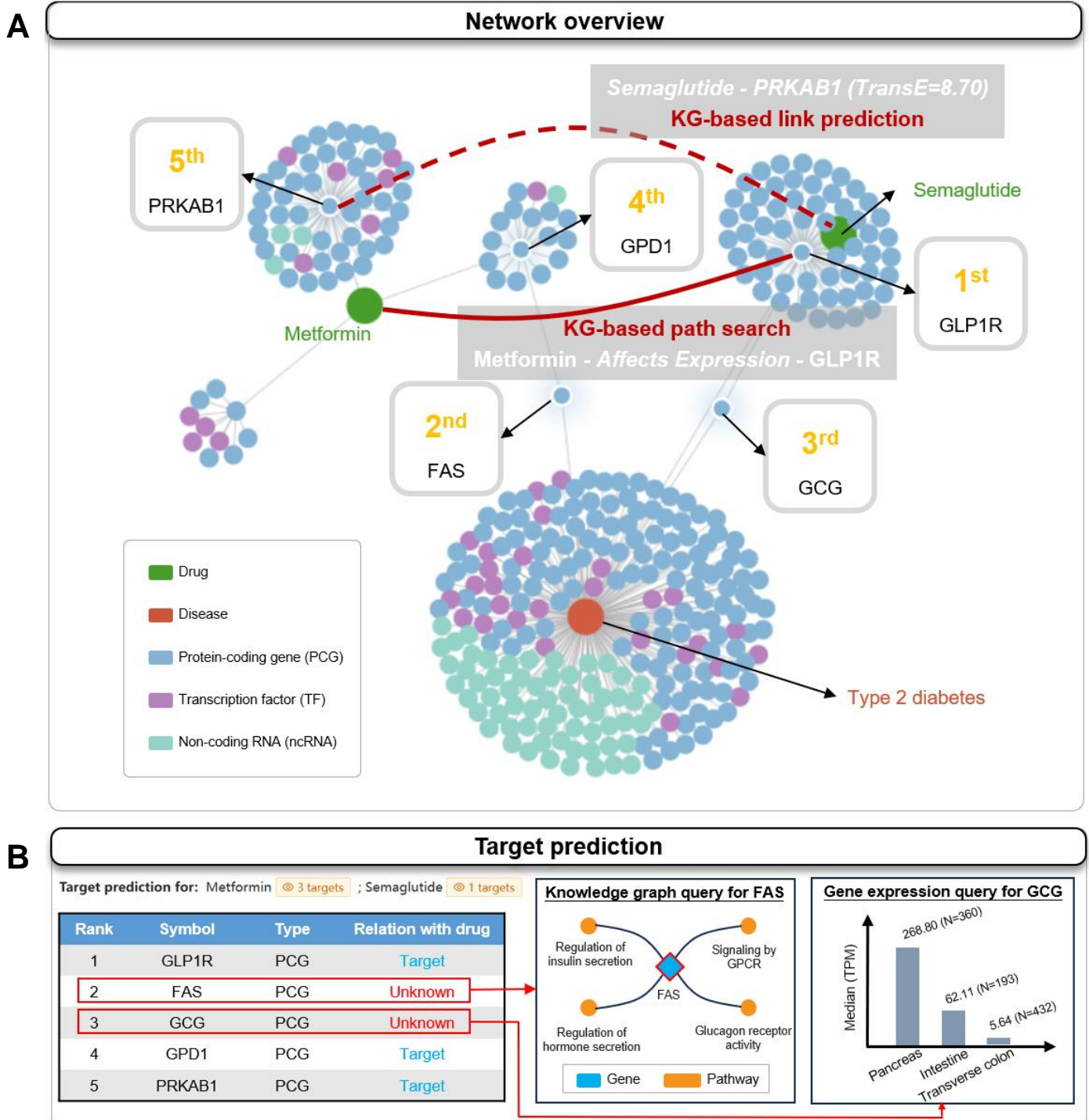

Figure 7

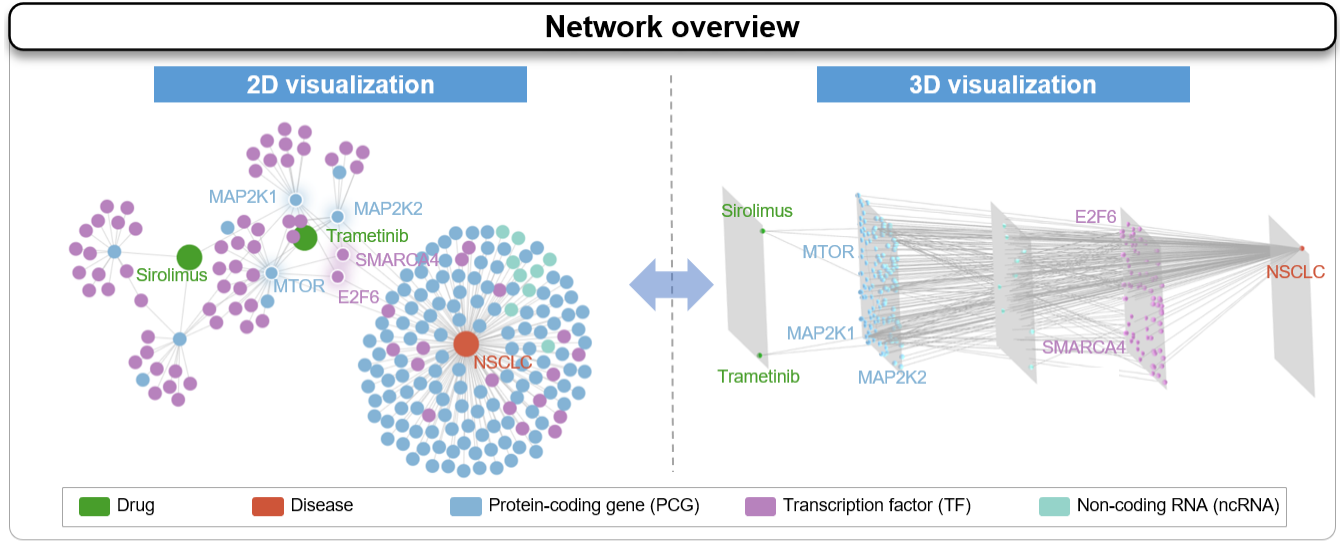

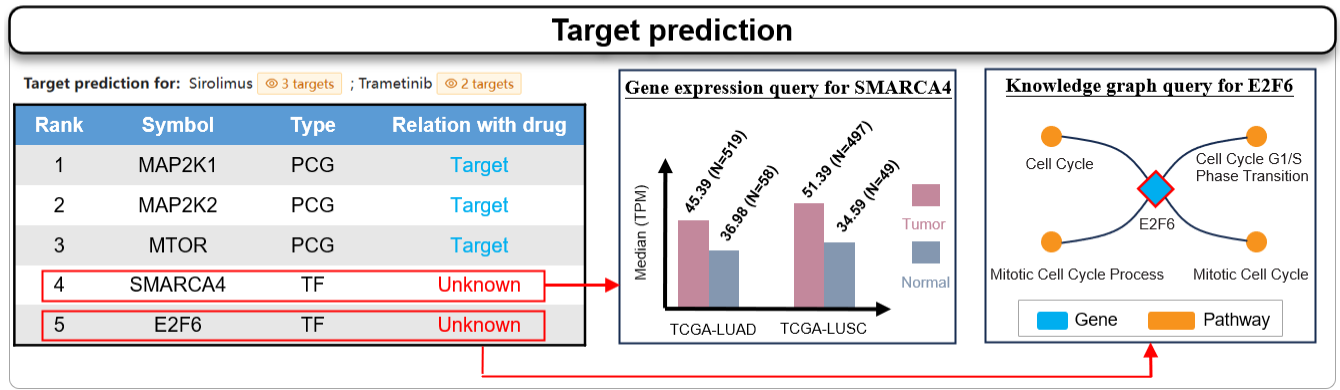

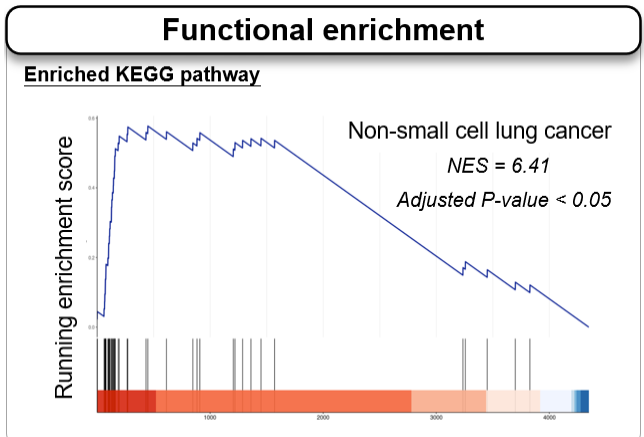

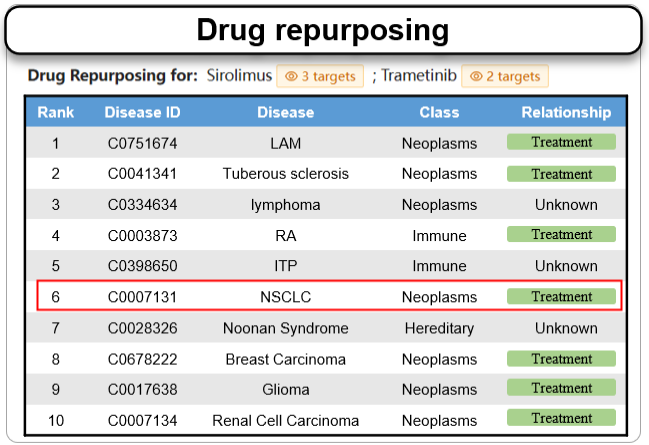

# Supplementary Figures

**POINT: a web-based platform for pharmacological investigation enhanced by multi-omics networks and knowledge graphs**

## Table of contents

**Supplementary Figure S1 | Quality control of multi-omics biological networks**

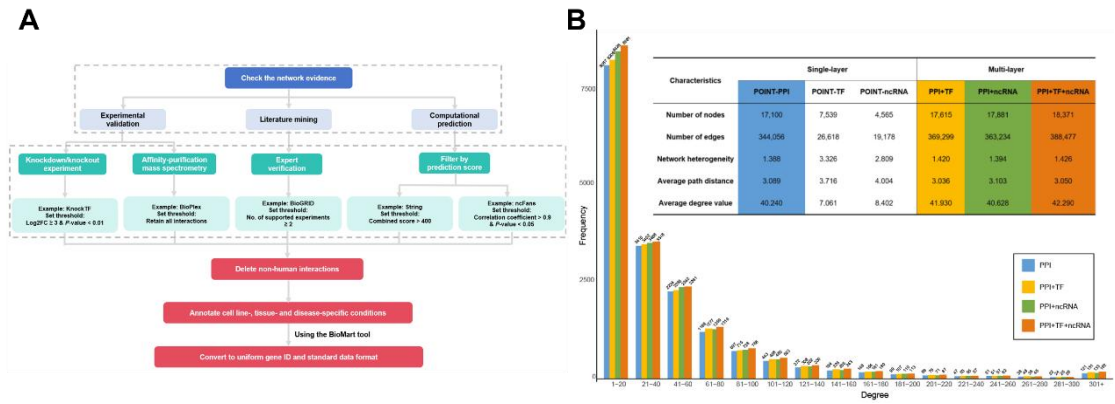

Quality control of multi-omics biological networks. (A) Quality control steps for databases containing single-layer networks of different omics. (B) Comparison of the pre-integrated multi-layer human network and its subnetworks.

**Supplementary Figure S2 | Network-based algorithms and tools provided in POINT**

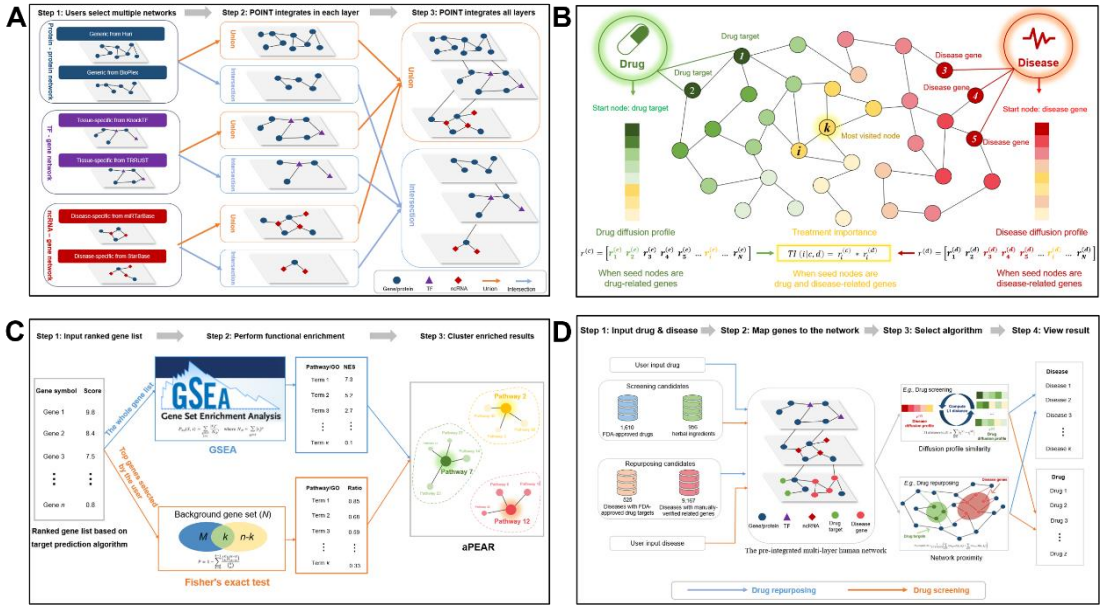

Network-based algorithms and tools provided in POINT. (A) Network integration strategies, including union and intersection. (B) Target prediction algorithm: random walk with restart (RWR). (C) Functional enrichment pipeline for predicted targets. (D) Drug screening and drug repurposing pipeline for inferring novel drug–disease relationships.

**Supplementary Figure S3 | Quality control steps for knowledge graphs**

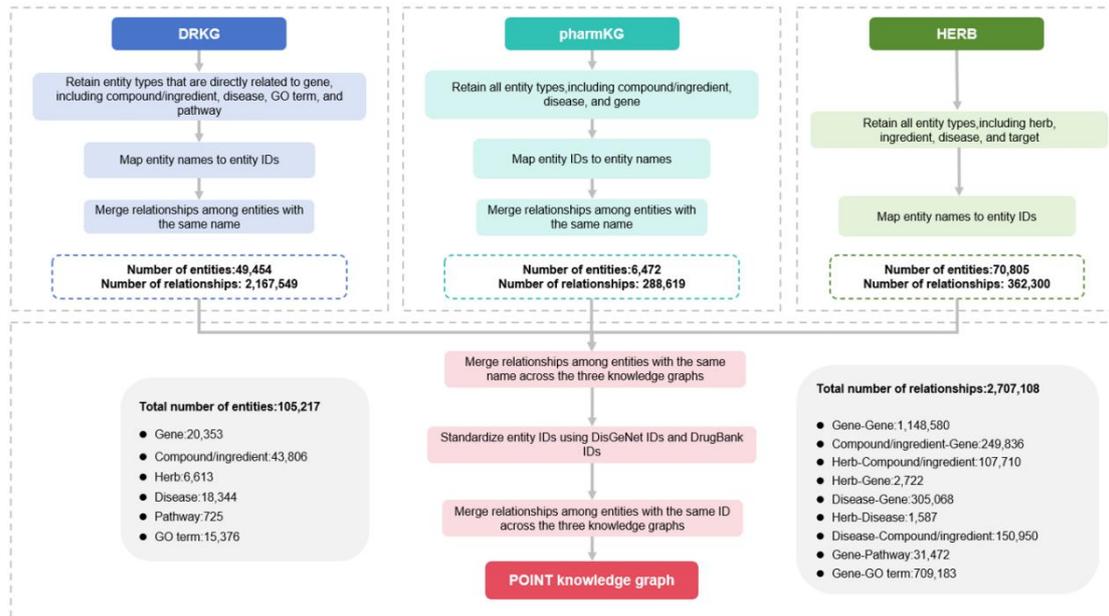

Quality control steps for knowledge graphs. The POINT knowledge graph was obtained by integrating three existing authoritative medical knowledge graphs: DRKG, PharmKG, and HERB.

**Supplementary Figure S4 | Knowledge graph-based algorithms provided in POINT**

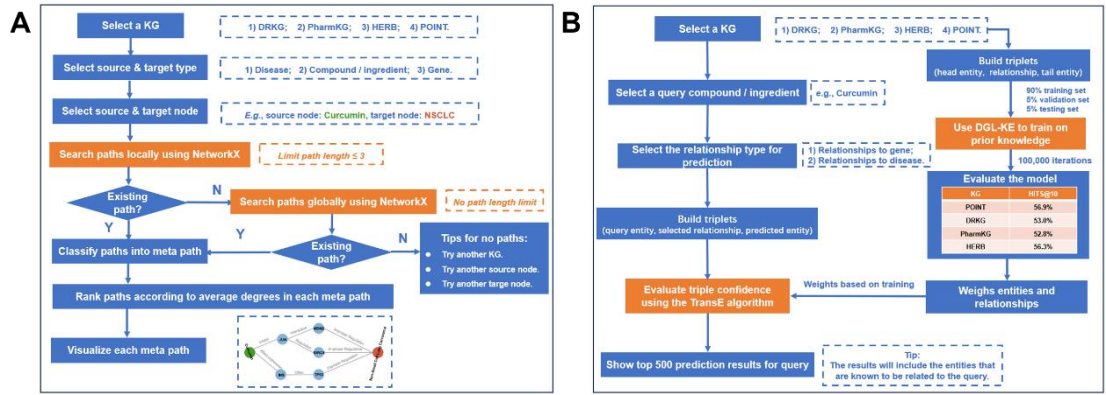

Knowledge graph-based algorithms provided in POINT. (A) Flowchart of the KG-based path search. (B) Flowchart of the KG-based link prediction.

**Supplementary Figure S5 | Influence of walk probability on target recall score**

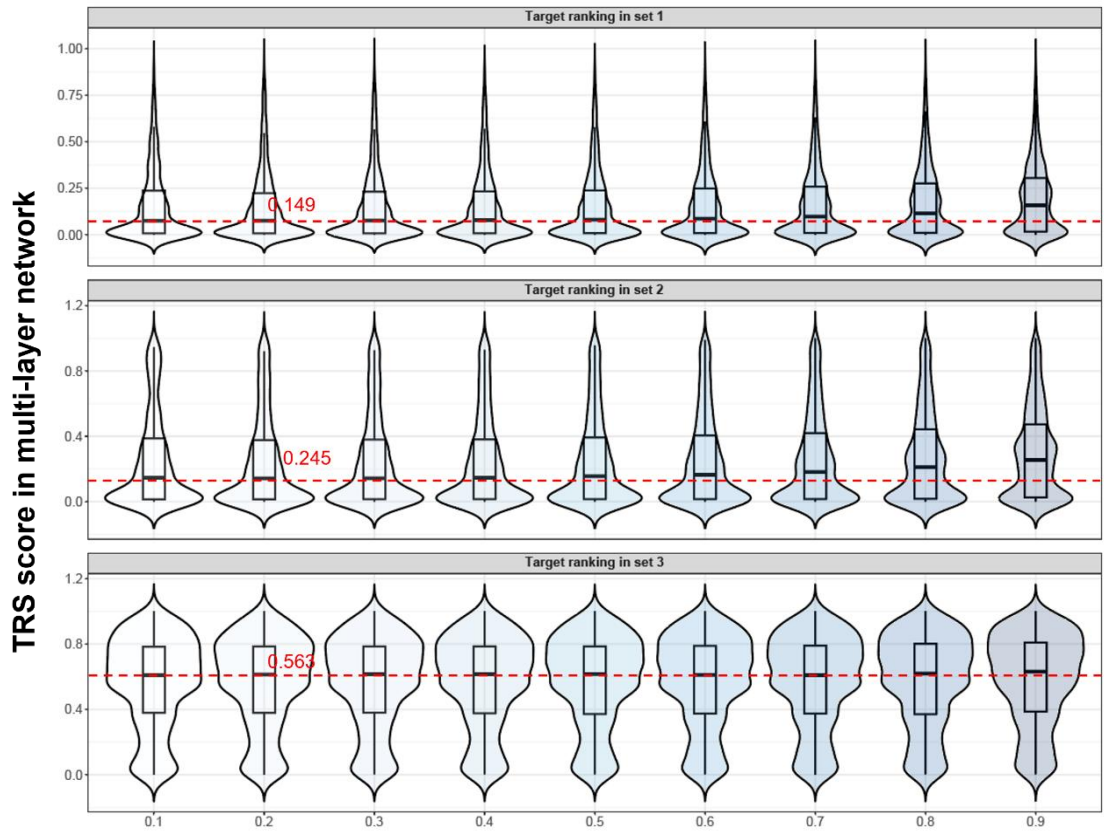

Target retrieval scores obtained using the RWR algorithm with varying walk probabilities (0.1-0.9) in pre-integrated multi-layer human network. The red dashed line indicates the average target retrieval score at a walk probability of 0.2.

**Supplementary Figure S6 | Proportion of network-based prediction results validated in KG**

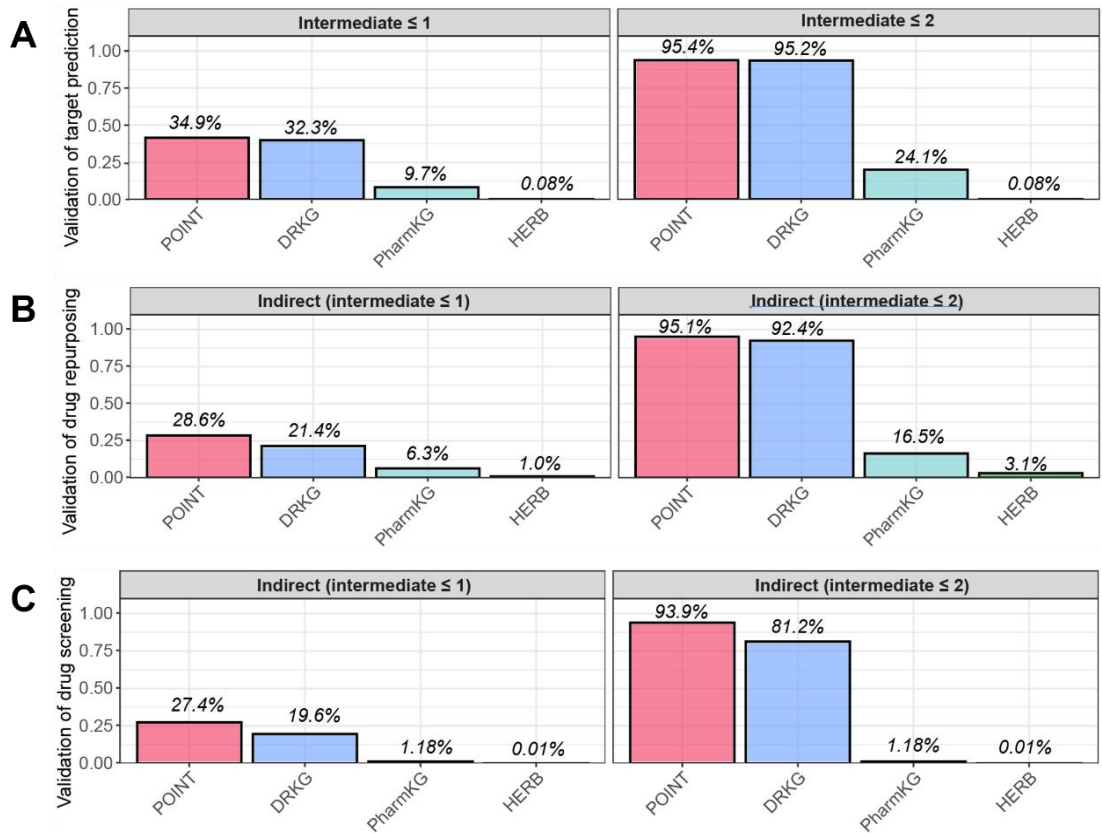

Percentage of network-based predictions validated in the KG provided by POINT. (A) Target prediction validation: proportion of top 500 drug-predicted target pairs for each drug successfully queried in different knowledge graphs. (B) Drug repurposing validation: proportion of top 500 drug-predicted disease pairs for each drug successfully queried in different knowledge graphs. (C) Drug screening validation: proportion of top 500 disease-predicted drug pairs for each disease successfully queried in different knowledge graphs.

**Supplementary Figure S7 | Input interface for knowledge graph-based analyses**

Input interface for knowledge graph-based analyses. (A) Input interface for knowledge graph-based path search. (B) Input interface for knowledge graph-based link prediction.

**Supplementary Figure S8 | Comparison of the performance of specific and generic networks to retrieve known targets**

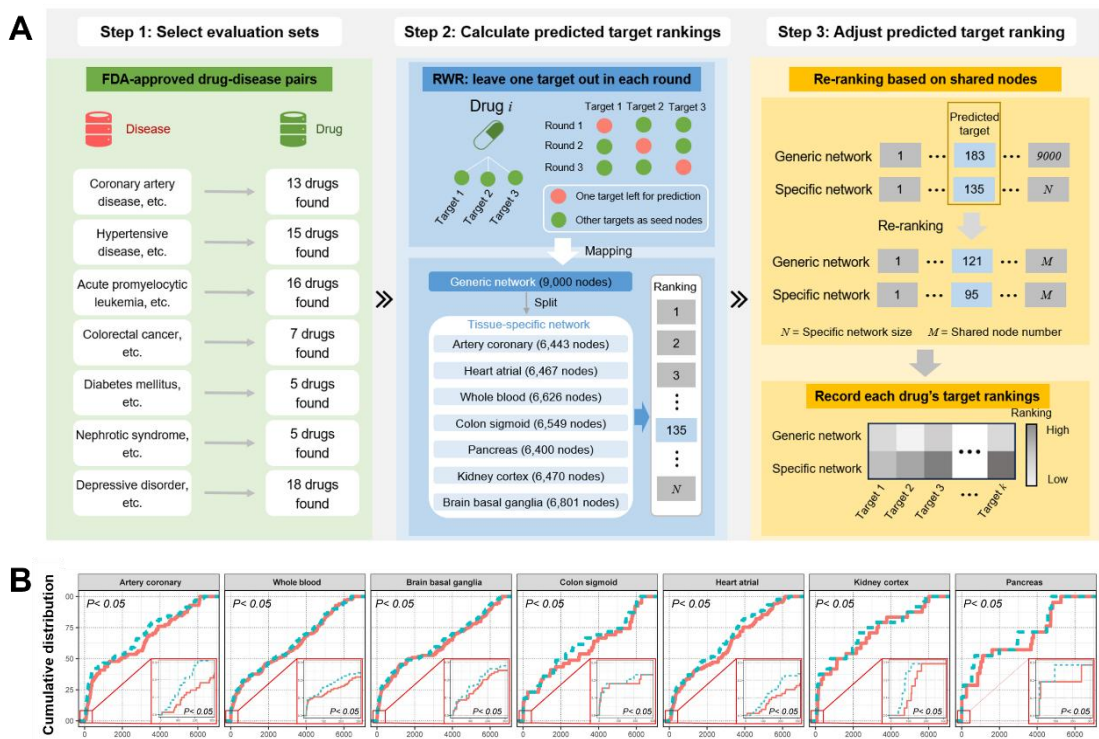

Comparison of the performance of specific and generic networks to retrieve known targets. (A) Evaluation scheme for comparing the performance of tissue-specific and generic networks in drug target prediction. Tissue-specific networks were constructed using HuRI yeast two-hybrid PPIs filtered by tissue-preferential expression (TiP score >0). (B) Cumulative distribution functions of predicted target rankings for drugs in the generic network and seven tissue-specific PPI networks. Rankings are shown for all shared nodes and the top 300 nodes, with cyan representing the specific network and red representing the generic network. Specific networks consistently outperformed generic networks.

**Table 1**

| POINT | Functionalities | Algorithms / Tools |
|---|---|---|
| **Network-based analysis** | Network integration | 1) Union; 2) Intersection. |
| | Network visualization (2D/3D) | 1) Echarts; 2) Arena3D. |
| | Target prediction | 1) Random walk with restart (RWR); 2) PageRank; 3) Degree. |
| | Functional enrichment | 1) GSEA; 2) Fisher's exact test; 3) aPEAR. |
| | Drug repurposing & screening | 1) Diffusion profile (DP) similarity; 2) Network proximity. |
| **KG-based analysis** | Path search | 1) NetworkX. |
| | Link prediction | 1) DGL-KE; 2) TransE. |